\documentclass[a4paper,11pt]{article}
\usepackage{jcappub} 

\usepackage{graphicx}	
\usepackage{amsmath}	
\usepackage{subfigure}
\usepackage{ifthen}
\usepackage{verbatim}
\usepackage{multicol}
\usepackage{pdflscape}
\newcommand{\wignerThree}[6]{
  \begin{pmatrix}
    #1 & #2 & #3 \\
    #4 & #5 & #6
  \end{pmatrix}
}

\arxivnumber{2403.16726} 
\title{Enhancing Primordial B-mode Detection: Comprehensive Delensing Pipelines for Improved Sensitivity to $r$}

\author{
Wen-Zheng Chen,$^{a,b}$
Yang Liu,$^{a}$
Siyu Li,$^{a}$
Bin Hu, $^{c,d}$
Hong Li$^{a}$
}
\affiliation{$^{a}$Key Laboratory of Particle Astrophysics, Institute of High Energy Physics, Chinese Academy of Sciences, Beijing 100049, People's Republic of China}
\affiliation{$^{b}$University of Chinese Academy of Sciences, Beijing 100049, People's Republic of China}
\affiliation{$^{c}$Institute for Frontier in Astronomy and Astrophysics, Beijing Normal University, Beijing, 102206, People's Republic of China}
\affiliation{$^{d}$Department of Astronomy, Beijing Normal University, Beijing 100875, People's Republic of China}

\emailAdd{liuy92@ihep.ac.cn}
\emailAdd{hongli@ihep.ac.cn}

\abstract{Recognizing the impact of contamination from weak gravitational lensing B-modes induced by Large Scale Structure, we examine delensing methods to enhance sensitivity to the tensor-to-scalar ratio $r$ in primordial B-mode detection experiments. This study presents a realistic pipeline to improve $r$ constraints using foreground-cleaned maps with negligible residuals. The pipeline, based on simulations, is adaptable for future experiments.
We focus on two delensing approaches: (1) subtracting the gradient-order lensing B-mode template, computed by convolving the E-mode with the lensing potential, from the observed B-mode signal; and (2) remapping observations using the estimated inverse deflection angle. For parameter constraints, we employ three models to reduce $r$ uncertainty and bias, finding consistent uncertainties across models, though biases vary due to the multipole-dependence of the delensing fraction.
We demonstrated this pipeline using simulated observation maps from future CMB polarization experiments, which included current representative ground-based small aperture telescopes (sub-1m), next-generation ground-based large aperture telescopes (6m), and highly competitive future space-based medium aperture missions (3m). Results show a delensing efficiency of 40\% with the small-aperture telescope alone, increasing to 65\% when combined with the large-aperture telescope, and 80\% with the satellite mission. These lead to reductions in $r$ uncertainty by 46\% for ground-based and 63\% for space missions. The most promising method adds the lensing template B-mode as an additional frequency channel, minimizing bias on $r$.}

\begin{document}
\maketitle
\flushbottom

\section{Introduction}

The Cosmic Microwave Background (CMB) radiation, originating roughly 380,000 years after the Big Bang, holds invaluable insights into cosmic evolution and serves as a pivotal probe for understanding the early universe. As CMB photons travel freely through space towards us from the last scattering surface, they are subject to gravitational deflection by the large-scale structures (LSS) scattered across the universe. This gravitational effect leads to distortions in the observed pattern of CMB anisotropies. This phenomenon is referred to as the CMB lensing effect, with a characteristic deflection angle typically around $2$ arcminutes.

Weak gravitational lensing, which is a secondary anisotropy on the CMB, has attracted lots of interest in recent years. Measurement of the CMB lensing provides a rare opportunity to obtain information on the distribution of cosmic gravitational field, so as to probe the accelerated expansion on large scales, to measure the dark energy equation of state, and to determine the total neutrino mass.
The first detection of CMB lensing was reported by \cite{smith2007detection}, achieving a significance of $3.4 \sigma$ by applying quadratic estimator techniques to all-sky maps from the WMAP satellite and correlating the results with radio galaxy counts from the NRAO VLA Sky Survey (NVSS) \cite{condon1998nrao}. Since then, many measurements of the lensing potential have been extensively studied in the literature\cite{namikawa2012full}, \cite{namikawa2014lensing}, \cite{story2015measurement}, \cite{Han:2023gvr}, \cite{Liu:2022beb}. 
With a minimum variance estimator derived from maximizing a posterior \cite{carron2017maximum}, Planck 2018 temperature and polarization maps improve the measurement of lensing potential to a confidence of $40 \sigma$\cite{aghanim2020planck_lensing}, and by using the lensing likelihood alone, the combined parameter of matter, $\sigma_8\Omega_m$ can be measured to an accuracy of a few percent, which is comparable to the constraints from weak lensing of galaxies. It is worth pointing out that there is a complementarity between the lensing statistics from low-redshift galaxies and those from high-redshift CMB, and thus their combination will be more helpful in improving the parameter measurements.

Weak gravitational lensing leads to a mixture of the polarization of CMB photon, converting a portion of the primary E-mode into a B-mode. The extra B-mode generated by weak gravitational lensing contaminates the measurements of primordial gravitational waves, so the lensing B-mode is a source of intrinsic noise for primordial B-mode detection \cite{manzotti2017cmb}. 
It is therefore essential to mitigate the lensing B-mode from CMB observation, and a straightforward way is to obtain a specific estimate of the lensing B-mode from the observed map, and then subtract it, a process known as "delensing". Delensing procedure becomes critical when the instrumental noise becomes subdominant compared to the lensed B-mode, which exhibits a nearly white spectrum at multipoles $\ell<1000$, corresponding to $\Delta_P \sim 5 \mu{\rm K}\text{-}{\rm arcmin}$.

The removal of lensing contamination from CMB anisotropy observation maps have received much attention and it has been widely studied in the literature \cite{kesden2002separation}, \cite{smith2012delensing}, \cite{sherwin2015delensing}, \cite{simard2015prospects}. The first direct delensing of CMB temperature anisotropies was achieved by using the cosmic infrared background (CIB) from star-forming dusty galaxies, employing a linear combination of the 545 GHz and 857 GHz maps as the CMB lensing tracer \cite{larsen2016demonstration}. This external delensing method led to a sharpening of the acoustic peaks in the temperature power spectrum, which was detected with a significance of $16\sigma$, indicating a successful delensing effort. Planck conducted internal delensing by using its own CMB temperature and polarization measurements to estimate the CMB lensing potential, and the estimated deflection map was then applied to partially reverse the lensing effects, as outlined in\cite{carron2017internal}. This process achieved delensing efficiencies of $29\%$ (TT),  $25\%$ (TE) and  $22\%$ (EE) of the power spectra, resulting in a $7\%$ reduction in lensing power in the B-mode map. A $22\%$ reduction in B-mode power variance was achieved using iterative delensing methods applied to deep polarization data, as described in \cite{adachi2020internal}, and the cosmological parameters from both lensed and delensed spectra are studied in \cite{han2021atacama}. 
The BICEP/Keck and SPTpol collaborations performed a joint analysis to improve constraints on $r$ by incorporating a lensing template into their likelihood function in 2016 \cite{ade2016improved}. In 2021, they achieved approximately a $10\%$ reduction in the uncertainty of $r$ by adding a lensing B-mode template to their analysis framework \cite{ade2021demonstration}.
In addition, the delensing effect has been extensively investigated in the literature to achieve greater delense efficiency\cite{namikawa2022simons}, with forecast of Simons observatory\cite{hertig2024simons}, CMB-S4\cite{belkner2023cmb}\cite{millea2020sampling}.


While delensing has been studied in the literature, developing an efficient and comprehensive measurement pipeline including delensing procedure as well as $r$ constraint from these existing delensing methods, would be highly valuable for detecting primordial gravitational waves (PGWs). In this paper, we explore two distinct delensing methods to evaluate the effectiveness of delensing in CMB polarization experiments across three different sensitivity levels, and we examine their impact on the parameter space relevant for measuring PGWs. The simulated observation maps used in this study are based on scans from Tibet, a mid-latitude site in the Northern Hemisphere, making them practically significant for the Ali CMB polarization experiment currently being constructed\cite{Li:2017drr,Li:2018rwc}. Additionally, this research provides valuable insights for future space-based missions with extremely high sensitivity. We present a comprehensive measurement pipeline for the tensor-to-scalar ratio, with the aim of identifying optimal combinations of methods for the future detection of PGWs.



The two methods—inverse-lensing and gradient-order methods—are employed to construct lensing templates, alongside three distinct delensing models (\textbf{L1}, \textbf{L2}, and \textbf{L3}), to improve constraints on the tensor-to-scalar ratio $r$ for future CMB experiments.
Our findings indicate that the two lensing template construction methods yield comparable results; however, the gradient-order method runs faster and consumes fewer resources. Among the delensing models, the cross-spectrum delensing model (\textbf{L3}) produces the smallest biases while maintaining comparable uncertainty on $r$. Additionally, we conduct a detailed study of the biases introduced by delensing procedure at the power spectra level. As this paper is focused on the delensing procedure, the whole pipeline is based on the foreground cleaned maps whose foreground residuals are negligible. \footnote{
We conducted a preliminary check on delensing with diffuse foreground remaining and found that the foreground primarily affects large scales. This influence can be modeled without significant modifications to our current work. A detailed discussion of its effects will be presented in our subsequent work.}

The outline of the paper is as follows: In Section \ref{sec:methods}, we introduce two delensing methods utilized in our work, followed by a comprehensive analysis of the biases introduced by delensing. Section \ref{sec:sims} provides details of the simulation setup for both the CMB maps and the lensing potential maps. Our main delensing results are presented in Section \ref{sec:result}, alongside a subsequent analysis involving noise debiasing and a comparison of the two delensing methods. 
We then constrain the tensor-to-scalar ratio, $r$, either by calculating a likelihood using the delensed BB power spectrum or by extending the likelihood function to include all auto- and cross-spectra between the lensing B-mode template and the observed B-mode.
Section \ref{sec:conclusion} concludes the paper. Additionally, in Appendix \ref{sec:algorithms}, we derive and prove some algorithms related to delensing expressions. In Appendix \ref{sec:filter}, we provide the derivation of the Wiener filter, which is widely employed in delensing procedures, along with an explanation of its significance in the delensing process. We explicitly describe how to add the lensing template to our parameter constraint pipeline in Appendix \ref{app:l3}. The effects induced by the tensor and scalar fluctuation in the delensing procedure are shown in Appendix \ref{app:component}.

\section{Delensing Methods}\label{sec:methods}
\label{sec:maths} 

Utilizing the projected gravitational potential, which gives rise to the CMB lensing effect, alongside the observed CMB polarization map, enables us to undertake the delensing. Two main methods are considered in this context: the first involves constructing a gradient-order B-mode template and subtracting it from the observed B-mode, and the second entails employing an inverse-lensing approach. In the latter method, delensing is accomplished by remapping the lensed maps using the inverse lensing deflection angle generated by the gravitational potential.

The temperature and polarization anisotropies of lensed CMB are described as:

\begin{equation}\label{EQ1}
    \begin{aligned}
        \tilde{\Theta}(\hat{n}) &= \Theta(\hat{n} + \alpha(\hat{n})) \\
        &= \Theta(\hat{n}) + \nabla_i\Theta(\hat{n})\alpha^i(\hat{n}) + \mathcal{O}(\phi^2),
    \end{aligned}
\end{equation}
\begin{equation}\label{EQ2}
    \begin{aligned}
		[\tilde Q\pm i\tilde U](\hat{n}) &= [Q\pm iU](\hat{n} + \alpha(\hat{n})) \\
			&= [Q\pm iU](\hat{n}) + \nabla_i[Q\pm iU](\hat{n})\alpha^i(\hat{n}) + \mathcal{O}(\phi^2),
    \end{aligned}
\end{equation}
where tildes denote the lensed fields. We exclusively focus on gradient modes, and express the deflection angle 
$\alpha(\hat{n})$ as the gradient of the lensing potential $\alpha(\hat{n}) = \nabla \phi(\hat{n})$  \cite{namikawa2014lensing}.

In harmonic space, the E- and B-modes can be defined using the spin-2 spherical harmonics $_2Y_{\ell m}$: 
\begin{equation}
	\begin{gathered}
		[E_{\ell m} \pm iB_{\ell m}] =  -\int d \hat{n} \ _2{Y}_{\ell m}^*(\hat{n}) \ [Q \pm iU](\hat{n}),
	\end{gathered}
\end{equation}
and the harmonic coefficients of lensing potential $\phi$ are defined as
\begin{equation}
	\begin{gathered}
		\phi_{\ell m} =  \int d \hat{n} \ {Y_{\ell m}^*}(\hat{n}) \ \phi(\hat{n}).
	\end{gathered}
\end{equation}

From Eq.(\ref{EQ2}), the lensed E and B-modes can be written as \cite{namikawa2014lensing}
\begin{equation}
	\begin{aligned}
		\tilde{E}_{\ell m} &= E_{\ell m} \\
  & \ +  \sum_{\ell' m'}\sum_{LM} \wignerThree{\ell}{\ell'}{L}{m}{m'}{M} \phi^*_{LM} \left\{\mathcal{S}^{(+)}_{\ell\ell'L} E^*_{\ell'm'} +  i\mathcal{S}^{(-)}_{\ell\ell'L}B^*_{\ell'm'}\right\},
	\end{aligned}
\end{equation}

\begin{equation}\label{EQ6}
	\begin{aligned}
		\tilde{B}_{\ell m} &= B_{\ell m} \\
  & \ +  \sum_{\ell' m'}\sum_{LM} \wignerThree{\ell}{\ell'}{L}{m}{m'}{M} \phi^*_{LM} \left\{\mathcal{S}^{(+)}_{\ell\ell'L} B^*_{\ell'm'} -  i\mathcal{S}^{(-)}_{\ell\ell'L}E^*_{\ell'm'}\right\},
	\end{aligned}
\end{equation}
where $\mathcal{S}^{(\pm)}_{\ell\ell'L}$ is given by
\begin{equation}
	\begin{aligned}
		\mathcal{S}^{(\pm)}_{\ell\ell'L} &= \frac{1 \pm (-1)^{\ell + \ell' + L}}{2}  \sqrt{\frac{(2\ell+1)(2\ell'+1)(2L+1)}{16\pi}} \\ & \ \times [-(\ell(\ell+1)+\ell'(\ell'+1)+L(L+1))] \wignerThree{\ell}{\ell'}{L}{2}{-2}{0}. 
	\end{aligned}
\end{equation}

\subsection{Reconstruct the gradient-order lensing B mode template}\label{sec:temp}

Starting from the second line of Eq.(\ref{EQ2}), we see that the gradient order lensing template can be written as $\nabla_i[Q\pm iU](\hat{n})\nabla^i\phi(\hat{n})$,
the gradient of QU maps and the lensing potential are given by
\begin{equation}
	\begin{gathered}
		\nabla \phi = -\frac{1}{\sqrt{2}} \left\{ \sharp \phi \bar m + \flat \phi m \right\},
	\end{gathered}
\end{equation}

\begin{equation}
	\begin{gathered}
		\nabla (Q + iU) = -\frac{1}{\sqrt{2}} \left\{ \sharp (Q + iU) \bar m + \flat (Q + iU) m \right\},
	\end{gathered}
\end{equation}
where we have used the relation of gradient operator defined by \cite{Okamoto:2003zw}:
\begin{equation}
	\begin{gathered}
		D_i [{_s}f(\hat n)] = -\frac{1}{\sqrt{2}} \left\{ \sharp {_s}f(\hat n) \bar m + \flat {_s}f(\hat n) m \right\},
	\end{gathered}
\end{equation}
where $\sharp$ and $\flat$ are the ladder operators, $D_i$ is the gradient operator, ${_s}f(\hat n)$ is a spin-s function and complex-conjugated vectors $\bar m$ and $m$ have the property 
\begin{equation}
	\begin{gathered}
		m \cdot m = \bar m \cdot \bar m = 0, \quad
		m \cdot \bar m =1,
	\end{gathered}
\end{equation}

Given QU maps and a potential map, our initial step involves decomposing them into harmonic coefficients of E- and B-modes. We prioritize E-modes due to their significantly higher strength compared to B-modes, thus neglecting any leakage from B-modes to E-modes. Subsequently, we transfer E-modes into a spin-1 field and a spin-3 field using ladder operators. Additionally, we transform the potential map into spin-1 and spin-$-1$ fields through the operation as follows:
\begin{equation}\label{EQ12}
	\begin{gathered}
        Q^3+iU^3 = \sharp (Q+iU) = \sum_{\ell m} \sqrt{(\ell+3)(\ell-2)} E_{\ell m} {_3}Y_{\ell m},
	\end{gathered}
\end{equation}
\begin{equation}\label{EQ13}
	\begin{gathered}
     Q^1+iU^1 = \flat (Q+iU) = -\sum_{\ell m} \sqrt{(\ell+2)(\ell-1)} E_{\ell m} {_1}Y_{\ell m},
	\end{gathered}
\end{equation}
\begin{equation}
	\begin{gathered}
    \sharp \phi  = \sum_{\ell m} \sqrt{\ell(\ell+1)} \phi_{\ell m} { _1}Y_{\ell m}, 
	\end{gathered}
\end{equation}
\begin{equation}
	\begin{gathered}
  \flat \phi  = -\sum_{\ell m} \sqrt{\ell(\ell+1)} \phi_{\ell m} {_{-1}}Y_{\ell m},
	\end{gathered}
\end{equation}
where $Q$ represents the real part of a field, and $U$ represents the imaginary part of a field. The index on top-right represent the spin of a field.

No B-mode should be included in Eq.(\ref{EQ12}) and Eq.(\ref{EQ13}), as introducing B-mode would intuitively imply that its presence will undergo lensing to the same order as the lensing B-mode (as Eq.(\ref{EQ6})). Following this consideration, we proceed to calculate the gradient order lensing template as follows:
\begin{equation}
	\begin{aligned}
		\nabla_i&[Q + iU](\hat{n})\nabla^i\phi(\hat{n}) \\
  &= \frac{1}{2} \left\{ \sharp \phi \bar m + \flat \phi m \right\}\left\{ \sharp (Q + iU) \bar m + \flat (Q + iU) m \right\} \\
			&= \frac{1}{2} [(\alpha^{1,R}+i\alpha^{1,I})(Q^1+iU^1)+(\alpha^{1,R}-i\alpha^{1,I})(Q^3+iU^3)],
	\end{aligned}
\end{equation}
where $\alpha^{R}$ and $\alpha^{I}$ represent the real and imaginary parts of the deflection field from lensing potential by spin raising. Separate the real and imaginary part of the lensing template field,
\begin{equation}\label{EQ17}
	\begin{gathered}
		Q = \frac{1}{2} [\alpha^{1,R}(Q^1+Q^3) - \alpha^{1,I}(U^1-U^3)],
	\end{gathered}
\end{equation}

\begin{equation}\label{EQ18}
	\begin{gathered}
		U = \frac{1}{2} [\alpha^{1,R}(U^1+U^3) + \alpha^{1,I}(Q^1-Q^3)].
	\end{gathered}
\end{equation}

The lensing B-mode template can be derived from the lensing template QU maps, and the delensed B-mode is obtained by subtracting the template map from the observed B-mode map. It is essential to note that the well-known E-to-B leakage must be carefully corrected when converting the Q and U Stokes parameters to E- and B-modes on a partial sky. To address this issue, we employ the ‘E-mode recycling’ method introduced by \cite{liu2019methods}.
In practice, we utilize the lensed QU maps rather than the unlensed ones because the former facilitates a cancellation between the high-order terms in the delensed power spectrum, whereas the latter does not, as described in \cite{BaleatoLizancos:2020jby}.

\subsection{Inverse-lensing method}

The second delensing method is more straightforward. From Eq.(\ref{EQ1}) and Eq.(\ref{EQ2}), we understand that the lensing effect can be reversed by remapping the observed photons back to their original positions. Therefore, the original unlensed field can be recovered by remapping the lensed field with an inverse deflection angle. The inverse deflection angle can be well defined because the points are remapped onto themselves after being deflected back and forth, as detailed in \cite{Diego-Palazuelos:2020lme}:

\begin{equation}\label{hat_n}
	\begin{gathered}
		\hat n + \beta(\hat n) + \alpha(\hat n+\beta(\hat n)) = \hat n,
	\end{gathered}
\end{equation}
where $\alpha(\hat n)$ and $\beta(\hat n)$ represent the deflection angle and the inverse deflection angle in the $\hat n$ direction on the surface of the sphere, and the unlensed field $X(\hat n)$ can be recovered from the lensed field $\tilde X(\hat n)$ by remapping the lensed field with an inverse deflection angle:
\begin{equation}
	\begin{gathered}
		\tilde X(\hat n) = X(\hat n+\alpha(\hat n)) \Leftrightarrow X(\hat n) = \tilde X(\hat n+\beta(\hat n)).
	\end{gathered}
\end{equation}
Eq.(\ref{hat_n}) can be solved by adopting a Newton-Raphson scheme, the inverse deflection angle $\beta(\hat n)$ can be iteratively calculated through \cite{Carron:2017vfg}
\begin{equation}\label{EQ21}    
	\begin{gathered}
		\beta_{(i+1)}(\hat n) = \beta_{(i)}(\hat n) - M^{-1}(\hat n + \beta_{(i)}(\hat n)) \cdot [\beta_{(i)}(\hat n) + \alpha(\hat n + \beta_{(i)}(\hat n)) ],
	\end{gathered}
\end{equation}
where $M(\hat n)$ is the magnification matrix, defined from the sphere's metric $g_{ab}$, as
\begin{equation}
	\begin{gathered}
		M_{ab}(\hat n) = g_{ab} + \nabla_a \alpha_b(\hat n).
	\end{gathered}
\end{equation}
We can then remap the observed map with the estimated inverse deflection angle in order to estimate the primary map.


\subsection{Delensing Bias}\label{sec:bias}
In this section, we will explore the biases introduced by the delensing procedure and suggest methods for their correction, known as debiasing.

Debiasing is necessary for obtaining an accurate measurement of $r$. 
We posit that the biases detailed below are distinct from the delensing biases stemming from mode overlap between the B-field to be delensed and the B-field utilized for reconstructing the lensing potential, as detailed in studies such as \cite{lizancos2021impact} and \cite{teng2011cosmic}. In these references, some bias terms were observed due to the overlap during the calculation of the power spectrum of the delensed B-mode, leading to a proven underestimation of the delensed B power.
In our subsequent derivation, we solely focus on separating each component contributing to the delensing residuals (biases), without considering the B-mode overlap mentioned above.

\subsubsection{Analysis of Biases at map level for the Gradient-Order Template Method}
The observed E and B modes with Wiener-filter (for a detailed description, please refer to Section \ref{subsec:mapproc}) are as follows:
\begin{equation}
	\begin{aligned}
		&E = \mathcal{W}^E(E^{lens}+E^{noise}), \\
		&B = \mathcal{W}^E(B^{lens}+B^{noise}), \\
		&\hat \phi = \mathcal{W}^{\phi}(\phi + \phi^{noise}),
    \end{aligned}
\end{equation}
with which we calculate the gradient order of the lensing B-modes template according to Section \ref{sec:temp}, following Eq.(\ref{EQA3}), and get:
\begin{equation}\label{EQ25}
	\begin{aligned}
    		B^{temp} &= \mathcal{B}^{(1)}[\mathcal{W}^E(E^{lens}+E^{noise}) \ast \mathcal{W}^{\phi}(\phi + \phi^{noise})] \\
		  &= \begin{aligned}[t] &\left\{\mathcal{B}^{(1)}[\mathcal{W}^E E^{lens} \ast \mathcal{W}^{\phi}\phi]\right\} + \mathcal{B}^{(1)}[\mathcal{W}^E E^{noise} \ast \mathcal{W}^{\phi}\phi] \\
		  	&+ \mathcal{B}^{(1)}[\mathcal{W}^E (E^{lens}+E^{noise}) \ast \mathcal{W}^{\phi}\phi^{noise}] 
		  \end{aligned} \\
		  &= B^{temp}_S + B^{temp}_N.
	\end{aligned}
\end{equation}

Here $\mathcal{B}^{(1)}[E \ast \phi]$ represents the operation of constructing the gradient order template with $E$ and $\phi$. 
We have checked that the tensor components have negligible contribution to the constructed lensing B-mode template, i.e. scalar E-mode is the primary contributor (see Appendix \ref{app:component}).
We artificially define the signal part ($B^{temp}_S$) which consists of the signal (the term in the brace), and the noise part ($B^{temp}_N$) introduced during the lensing template construction procedure.

Furthermore, since the input $E$ and $\hat \phi$ fields  are filtered, $B^{temp}_S$ is a filtered estimate of the lensing B-modes. In fact, a deviation is introduced by the Wiener filter, and certain high-order terms are neglected when constructing the template. Therefore, we express $B^{temp}_S$ as:

\begin{equation}
	\begin{gathered}
        B^{temp}_S = B^{lens} + N',
    \end{gathered}
\end{equation}
where $N'$ denotes the bias induced by both the Wiener filter and the method itself. 
This bias can be understood from Eq.(\ref{EQ6}), where the gradient-order template is a good approximation on large scales. However, with increasing $\ell$, the higher-order terms of the lensing B-modes become more and more significant. Therefore, the neglect of these high-order terms contributes to the bias, particularly on small scales. We denote $N'$ as the intrinsic bias to distinguish it from the noise part, as this term persists even when delensing with a noiseless map.

Then the delensed B-modes can be written as:
\begin{equation}\label{EQ26}
	\begin{aligned}
		B^{del} &= B^{obs} - B^{temp} \\
				&= (B^{tens}+B^{lens}+B^{noise}) - (B^{lens} + N' + B^{temp}_N) \\
				&= B^{tens}+B^{noise} - N' - B^{temp}_N \\
                &= B^{tens}+B^{noise} + B^{res} + B^{del}_{N}. \\ 
    \end{aligned}
\end{equation}

We have redefined two terms in the last line to align with an uniform format. Therefore, there are three components contributing to the delensed B-modes: the delensing residual $N'$ (also referred to as $B^{res}$ or $B^{del}_S$ hereafter), the intrinsic noise in observed B-modes $B^{noise}$, and $B^{temp}_N$, the noise introduced during the lensing template construction procedure. We denote the latter two as the delensing noise.

\subsubsection{Analysis of Biases at map level for the inverse-lensing method}

The inverse-remapped B-mode can be written as (check Eq.(\ref{EQA3})):
\begin{equation}\label{EQ27}
	\begin{aligned}
    		B^{del} &= \mathcal{B} [\mathcal{W}^E (B^{tens} + B^{lens} + B^{noise}) \star \beta ] \\
				& \approx  \mathcal{B} [\mathcal{W}^E (B^{tens}+ B^{lens} + B^{noise}) \star (\beta_\phi + \beta_{noise}) ] \\
				& \approx\left\{ \mathcal{B} [\mathcal{W}^E B^{tens} \star \beta ] \right\} + \left\{ \mathcal{B} [\mathcal{W}^E B^{lens} \star \beta_\phi ] \right\} \\
				&+ \left\{ \mathcal{B} [\mathcal{W}^E {B^{lens}} \star \beta_{noise} ] + \mathcal{B} [\mathcal{W}^E  B^{noise} \star \beta ]  - \mathcal{W}^E B^{lens} \right\} \\
				&= (\mathcal{W}^E)B^{tens} + B^{del}_S + B^{del}_{N} + (\mathcal{W}^E)B^{noise}. 
	\end{aligned}
\end{equation}
Here, $\mathcal{B}[B \star \phi]$ represents the delensing operation of lensing B mode with $\phi$ to obtain the delensed B-modes. 
We have checked that the inverse-lensing procedure have negligible effect on the tensor B-mode, so the first brace is almost equal to the filtered input tensor B-mode with the change accounted for by $B_S^{del}$ (see Appendix \ref{app:component}).
$B^{del}_S$ includes the delensed B signal (the second brace), which is the desired output and arises from the method itself, therefore exists even under noiseless observation. 
While the last brace contains both the intrinsic noise $(\mathcal{W}^E)B^{noise}$ and the noise part $B^{del}_{N}$ introduced by the delensing procedure.

The lensing template B-mode is then:
\begin{equation}\label{EQ28}
	\begin{aligned}
		B^{temp} &= B^{obs} - B^{del} \\
				&= [B^{tens}+B^{lens} + B^{noise}] - [B^{del}_S + B^{del}_N] \\
                 &- (\mathcal{W}^E)B^{tens} - (\mathcal{W}^E)B^{noise}  \\
				&\approx [B^{lens} - B^{del}_S] + [- B^{del}_N] \\
				&= B^{temp}_S + B^{temp}_N,
	\end{aligned}
\end{equation}
where $B^{temp}_S$ encompasses the lensing template B-mode we aim to obtain, and $B^{temp}_N$ represents the noise component. It's important to note that $B^{temp}_S$ still retains an intrinsic bias. In the third line, $B^{tens}$ and $B^{noise}$ vanishes as we assume the instrumental noise is relatively small compared to the E-modes signal, this is quite realistic for ongoing and future CMB experiments.
\begin{equation}\label{EQ29}
	\begin{aligned}
		B^{temp}_S = B^{lens} - B^{del}_S,
	\end{aligned}
\end{equation}
where the bias $B^{del}_S$ arises from the approximation introduced in obtaining the inverse deflection angle, and we denote it as the intrinsic bias to distinguish it from the noise bias.

\subsubsection{A summary of the bias } \label{sec:forms_of_bias}
To summarize above discussion, we hereby present the following conclusion:
\begin{equation}\label{eq:conclude_eq}
	\begin{aligned}
        &B^{temp} = B^{temp}_S + B^{temp}_N, \\
        &B^{del} = B^{tens} + B^{del}_S + B^{del}_N  + B^{noise},
	\end{aligned}
\end{equation}
where the terms on RHS are:
\begin{equation}\label{eq:conclude_eq_temp}
	\begin{aligned}
        B^{del}_S &= B^{res} = B^{lens} - B^{temp}_S, \\
        B^{del}_N &= - B^{temp}_N.
	\end{aligned}
\end{equation}

Therefore, the power spectrum of the lensing B-mode template can be written as,
\begin{equation}\label{eq:cl_temp_debias}
	\begin{aligned}
        C^{LT}_{\ell} &= \langle B^{temp}_{\ell m}B^{temp *}_{\ell m} \rangle  \\
        &= \langle B^{temp}_{S,\ell m}B^{temp *}_{S,\ell m} \rangle + 2\langle B^{temp}_{S,\ell m}B^{temp *}_{N,\ell m} \rangle + \langle B^{temp}_{N,\ell m}B^{temp *}_{N,\ell m} \rangle,
	\end{aligned}
\end{equation}
where the last two terms are regarded as template noise terms and can be estimated by simulation-based approach.
And the delensed power spectrum is simply
\begin{equation}\label{eq:cl_de_debias}
	\begin{aligned}
        C^{del}_{\ell} &= \langle B^{del}_{\ell m}B^{del *}_{\ell m} \rangle \\
                    &= \langle B^{tens}_{\ell m}B^{tens *}_{\ell m} \rangle
                    +\langle B^{del}_{S,\ell m}B^{del *}_{S,\ell m} \rangle 
                        + \langle B^{del}_{N,\ell m}B^{del *}_{N,\ell m} \rangle \\
                        & \ + \langle B^{noise}_{\ell m}B^{noise *}_{\ell m} \rangle 
                        + 2\langle B^{del}_{S,\ell m}B^{del *}_{N,\ell m} \rangle
                        + 2\langle B^{del}_{S,\ell m}B^{noise *}_{\ell m} \rangle \\
                        & \ + 2\langle B^{del}_{N,\ell m}B^{noise *}_{\ell m} \rangle
	\end{aligned}
\end{equation}

\begin{figure}
	\includegraphics[width=\columnwidth]{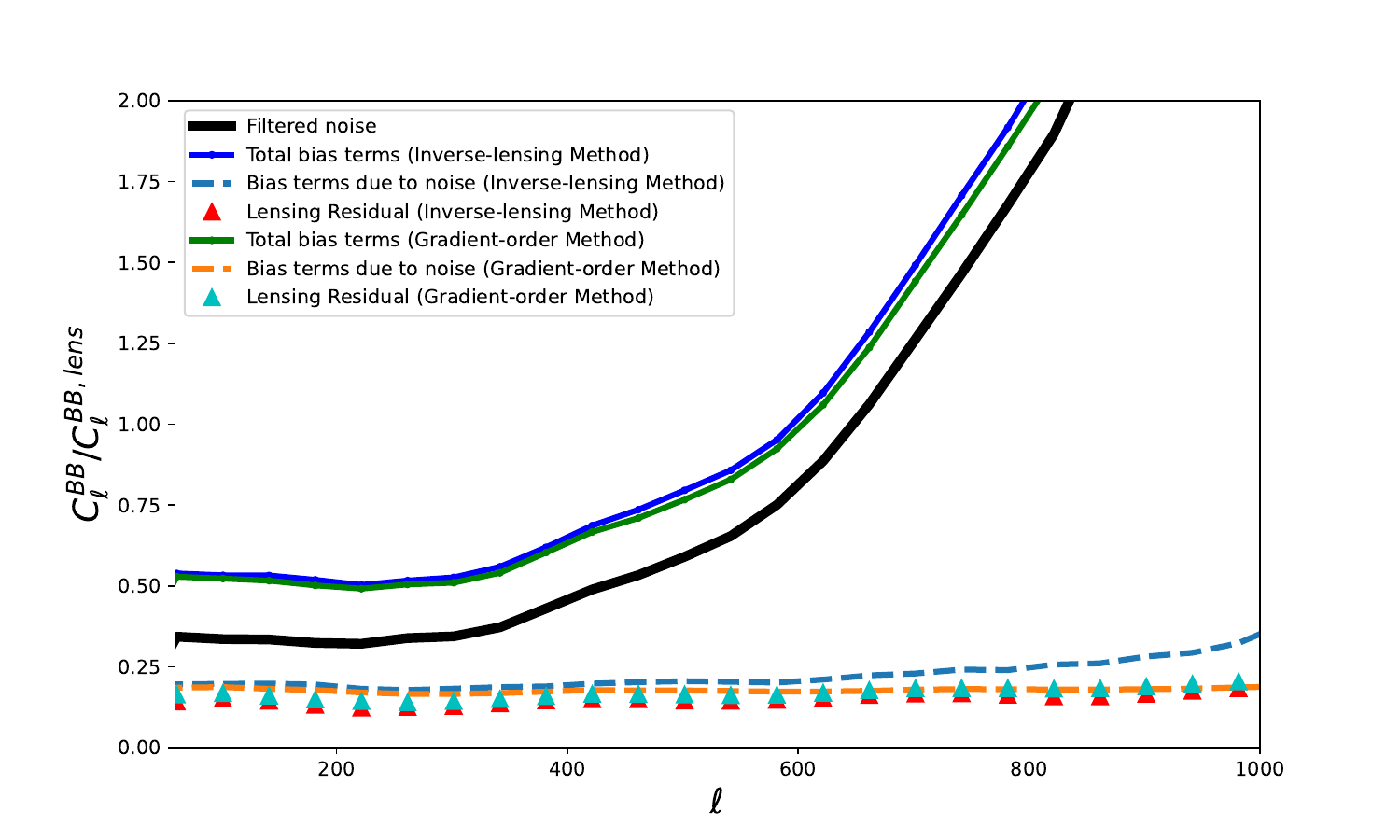}
	\caption{Terms of the delensed BB power divided by the theory lensed BB power.}
	\label{fig:terms}
\end{figure}

Notice that the $B^{tens}_{\ell}$ does not correlate with other terms.
We show the quantitative relationships of these terms in Fig.\ref{fig:terms}, averaged over 100 simulations. The chosen parameter settings are intended for the ground experiments (see Table \ref{tab:parameters}). The ‘lensing residual' denotes to $\langle B^{del}_{S,\ell m}B^{del *}_{S,\ell m} \rangle$, which quantifies the remaining lensing signal after delensing. The ‘Bias terms due to noise' represents the total of $\langle B^{del}_{N,\ell m}B^{del *}_{N,\ell m} \rangle$, $\langle B^{del}_{S,\ell m}B^{del *}_{N,\ell m} \rangle$, $\langle B^{del}_{S,\ell m}B^{noise *}_{\ell m} \rangle$ and $\langle B^{del}_{N,\ell m}B^{noise *}_{\ell m} \rangle$, which we believe quantifies the effects caused by the instrumental noise and reconstruction noise.  The ‘total bias terms' are the sum of the previous two. 
It is evident that the ‘lensing residuals' derived from the two methods are quite similar (with a difference of $\mathcal{O}(1)\%$), which represents the residual lensing signal power. 
However, although they exhibit similar performance, their mechanisms are different. 
The Gradient-order template method is defined to subtract the gradient-order lensing effect at the map level, with parts of the high-order terms eliminated when using the lensed QU fields as input, it encounters high-order cancellations when calculating the power spectrum on large scales. 
In contrast, the Inverse-lensing method inherently aims to eliminate the lensing effect by definition. 
Therefore, we believe the Inverse-lensing method is the optimal one with more accuracy on all scales by definition. 
Additionally, it is noteworthy that the ‘Bias terms due to noise' generated by the two methods are relatively consistent on large scales. However, the Inverse-lensing method produces larger ‘Bias terms due to noise' on small scales. We believe this is because, by its definition, it tends to absorb more noise into the higher-order terms.

We emphasize that although these two methods yield consistent forms of output, there are minor difference in certain terms due to higher-order effects. However, this does not lead to significant deviations in the final results at the scales of interest. We will examine these difference in the Section \ref{sec:main_result}.

We notice that in some literature (e.g. \cite{smith2012delensing}), the delensing residual is defined as $B^{res} = B^{lens} - B^{temp} = B^{del} - B^{noise}$, where $B^{res}$ represents the sum of our $B^{res}$ and $B^{temp}_N$. It should to clarify that this expression is merely an alternative formulation with equivalent meaning.

\subsubsection{The delensing effect on constraining $r$}\label{sec:summary_of_bias}
In a realistic observation case, the observed B-modes and delensed B-modes can be written as:
\begin{equation}\label{eq:bobs}
	\begin{aligned}
        B^{obs} &= B^{tens} + B^{lens} + B^{noise}, \\
        B^{del} &= B^{tens} + B^{res}  + B^{del}_{N} + B^{noise}.
	\end{aligned}
\end{equation}

Where $B^{del}$ represents the noisy delensed B-modes we construct, $B^{lens}$ is the lensing B-modes, $B^{tens}$ is the true unlensed B-modes from tensor fluctuations we seek, $B^{res}$ denotes the lensing residual field, $B^{del}_{N}$ denotes the noise introduced during the delensing procedure (delensing noise), and $B^{noise}$ represents the intrinsic noise of the observed lensed B-modes.

The BB angular power spectra without and with delensing can be written as:

\begin{equation}\label{eq:clobs}
	\begin{aligned}
		C_{\ell}^{obs} &= C_{\ell}^{tens} + C_{\ell}^{lens} + N_\ell^{BB},\\
            C_{\ell}^{del} &= C_{\ell}^{tens} + C_{\ell}^{res} + N_\ell^{del} + N_\ell^{BB}.
	\end{aligned}
\end{equation}

Here we define the delensing fraction as:
\begin{equation}\label{eq:delenfrac1}
	\begin{gathered}
		f_{dl} = (C_{\ell}^{del} - N_\ell^{BB}) / C_{\ell}^{lens},
	\end{gathered}
\end{equation}
this equates to :
\begin{equation}\label{eq:delenfrac2}
	\begin{gathered}
		f_{dl} = (C_{\ell}^{res} + N_\ell^{del}) / C_{\ell}^{lens}
	\end{gathered}
\end{equation}
when assuming $r=0$.
For cases without delensing, $f_{dl}=1$.
Then the BB angular power spectra without and with delensing can be simplified as:
\begin{equation}\label{eq:clobs1}
	\begin{gathered}
		C_{\ell} = C_{\ell}^{tens} + f_{dl}C_{\ell}^{lens} + N_\ell^{BB}.
	\end{gathered}
\end{equation}
When fitting the tensor-to-scalar ratio ($r$) using the angular power spectra, we often utilize the following parameterization:

\begin{equation}\label{eq:fitr}
	\begin{gathered}
		C_{\ell} = rC_{\ell}^{tens}(r=1) + A_LC_{\ell}^{len,th} + N_\ell^{BB}.
	\end{gathered}
\end{equation}
$C_{\ell}^{len,th}$ represents the lensing BB spectra from the theory, and $A_L$ is a constant characterizing the lensing amplitude. $N_\ell^{BB}$ is estimated from the observed B modes noise only simulations. When we do not perform the delensing operation, $A_L$ is approximately equal to 1, and the estimation of $r$ is unbiased. During delensing, if $f_{dl}$ is independent of $\ell$, the estimation of $r$ remains unbiased. Typically, $A_L<1$. However, if $f_{dl}$ heavily depends on $\ell$, we can adjust the parameterization to:

\begin{equation}\label{eq:fitr2}
	\begin{gathered}
		C_{\ell} = rC_{\ell}^{tens}(r=1) + A_LC_{\ell}^{len,th} + N_\ell^{del} + N_\ell^{BB}.
	\end{gathered}
\end{equation}
The estimation of $N_\ell^{del}$ is obtained from simulations.

The uncertainty of $r$ can be roughly expressed as:

\begin{equation}\label{eq:sigmar}
	\begin{gathered}
		\sigma(r=0) = \left[\sum_l\frac{(2l+1)f_{sky}}{2}\left(\frac{C_l^{tens}(r=1)}{f_{dl}C_l^{len}+N_l^{BB}}\right)^2\right]^{-1/2}.
	\end{gathered}
\end{equation}

From Eq.(\ref{eq:sigmar}), it can be observed that the uncertainty in $r$ decreases along with $f_{dl}$. This reduction in uncertainty is the motivation behind performing the delensing operation.

\section{Data simulation}\label{sec:sims}
 
We utilize simulated data to illustrate the feasibility of delensing methods. We simulate far future CMB polarization observations from various experiments, including one medium-aperture satellite mission of $3m$ telescope (sMAT), one small-aperture ground based  $80 cm$ telescope (gSAT), and one ground $6m$ large-aperture telescope (gLAT), to predict the delensing efficiency of different experiments and methods. This paper focuses on the study of delensing methods, and we do not consider the influence of foregrounds. In our next article, we will investigate the impact of foregrounds on delensing and its effect on the tensor-to-scalar ratio \( r \) for a more realistic simulation.

\begin{figure}
	\includegraphics[width=\columnwidth]{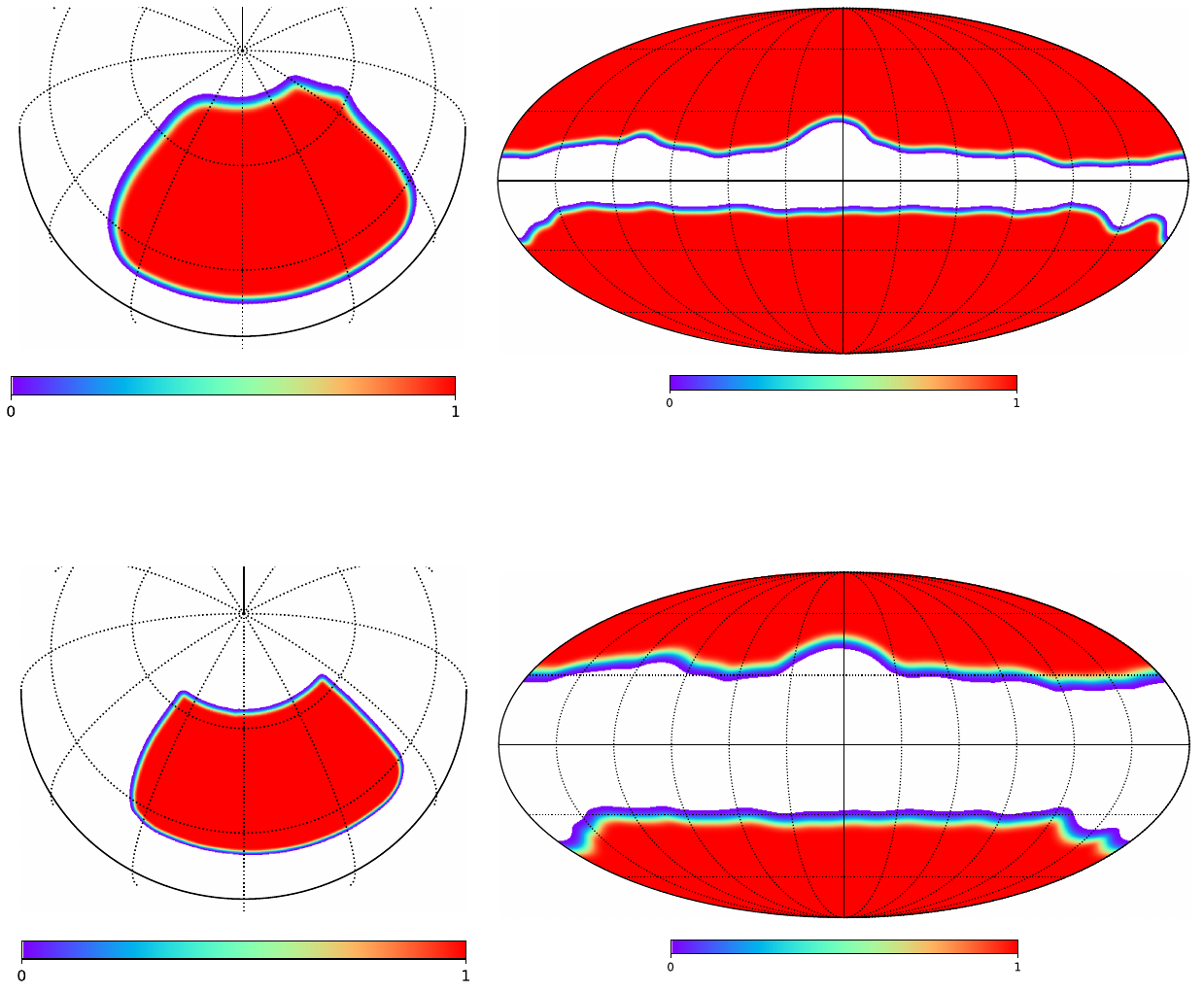}
	\caption{Various masks used in the simulation. Left panels are used for gSAT and gLAT and right panels are used for sMAT. Panels at the top are the apodized mask used for the delensing procedure,  and panels at the bottom show the apodized mask used for the calculation of pseudo-Cl with \texttt{NaMaster}.}
	\label{fig:mask}
\end{figure}

The input unlensed CMB maps are Gaussian realizations generated from a specific power spectrum obtained from the Boltzmann code \texttt{CAMB} \cite{lewis2011camb}, using the Planck 2018 best-fit cosmological parameters \cite{aghanim2020planck_parameter} with the tensor scalar ratio $r = 0$. We then utilize the \texttt{Lenspyx} package, which implements an algorithm to distort the primordial signal given a realization of the lensing potential map from $C_{\ell}^{\phi\phi}$. 
The lensed CMB maps are smoothed by the instrumental beam sizes listed in Table \ref{tab:parameters} respectively. All the maps are in the \texttt{HealPix} pixelization scheme at $\texttt{NSIDE}=2048$. The noise in the sky patch is homogeneous with noise levels listed in Table \ref{tab:parameters}. Finally, we add the smoothed lensed CMB maps to the noise maps, and then multiply them by the apodized sky patch  masks shown in Fig.\ref{fig:mask} to produce simulated observed maps for 3 experiments.

\begin{table}
	\centering
	\caption{The parameters of the three experiments.}
	\label{tab:parameters}
	\begin{tabular}{lcccr} 
		\hline
		Experiment & $\theta_{\text{FWHM}}$ & $\sigma_{\text{noise}}$ & $f_{\text{sky}}$ & $\ell \quad range$\\
		\hline
		gLAT & 1.4 arcmin & 6.0 $\mu{\rm K}\text{-}{\rm arcmin}$ & 0.1 & (300,5000) \\
		gSAT & 11.6 arcmin & 2.0 $\mu{\rm K}\text{-}{\rm arcmin}$ & 0.1 & (30,1000) \\
		sMAT & 2.8 arcmin & 1.5 $\mu{\rm K}\text{-}{\rm arcmin}$ & 0.8 & (2,3800)	  \\
		\hline
	\end{tabular}
\end{table}

\section{Delensing implementation and Results}\label{sec:result}
\begin{figure}
	\includegraphics[width=\columnwidth]{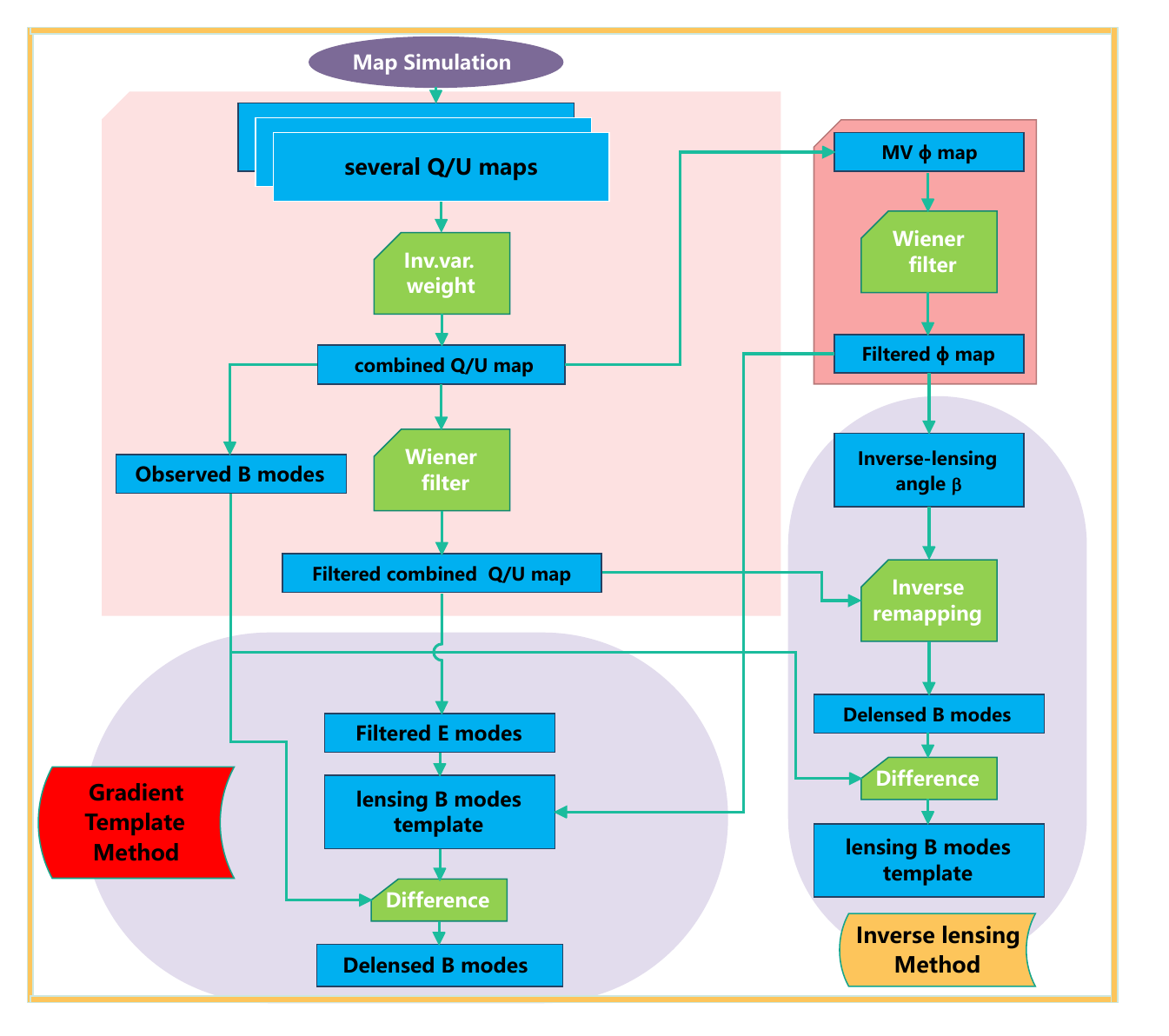}
	\caption{Flowchart of the delensing pipeline. We separate our pipeline into four parts: map simulation (top), lensed CMB maps processing (left upper), lensing potential processing (right upper) and the implementation of two delensing methods (lower).}
	\label{fig:pipeline}
\end{figure}

In this section, we first describe our implementation of delensing, then we present our main results of delensing and some further results after noise debiasing. Fig.\ref{fig:pipeline} shows our flowchart of the delensing pipeline.

\subsection{Lensed CMB maps processing}\label{subsec:mapproc}

We initially perform Wiener-filter locally on the observed $Q$ and $U$ in spherical-harmonic space to enhance the signal-to-noise ratio (SNR) prior to delensing \cite{Green:2016cjr}, as follows:
\begin{equation}
	\begin{gathered}
		Q_{\ell m} \Rightarrow \frac{C_{\ell}^{EE}}{C_{\ell}^{EE} + N_{\ell}^{EE}} Q_{\ell m}, \\
		U_{\ell m} \Rightarrow \frac{C_{\ell}^{EE}}{C_{\ell}^{EE} + N_{\ell}^{EE}} U_{\ell m}. \\
	\end{gathered}
\end{equation}

We provide a thorough derivation and clarification of the Wiener filter in the delensing operation in Appendix \ref{sec:filter}.


In our work, we first combine several observed maps with inverse-variance weight, and then perform Wiener-filter on the combined map. Notice that this diagonal filtering is sub-optimal for a cut sky case, and a more optimal filtered E-mode can be obtained by solving the following equation \cite{eriksen2004power}: 

\begin{equation}
	\begin{gathered}
		\left[ 1 + \sum_{m} \mathbf{C}^{1/2} {b}_{m} \, \mathbf{Y}^{\dagger} \mathbf{N}_{m}^{-1} \mathbf{Y} \, {b}_{m} \, \mathbf{C}^{1/2} \right] \left( \mathbf{C}^{-1/2} \mathbf{x}^{w} \right) = \sum_{m} \mathbf{C}^{1/2} {b}_{m} \, \mathbf{Y}^{\dagger} \mathbf{N}_{m}^{-1} \mathbf{d}_{m} \, .
	\end{gathered}
\end{equation}
where $m$ is index for all the input maps. The method will optimally combine several input observed maps (e.g. across several frequencies) with an effective filtering under the assumption that foreground is negligible. The harmonic coefficients of the Wiener-filtered E- and B-modes are obtained by questing for the vector $\mathbf{x}^{w}$, and the real-space vector $\mathbf{d}_{m}$ contains the Stokes $Q$ and $U$ maps. The matrix $\mathbf{C}$ is the diagonal signal covariance of the lensed E- and B-modes, and $\mathbf{C^{1/2}}$ is its square root. The matrix $\mathbf{N}_{m}$ is the covariance matrix of the instrumental noise of the input maps, and ${b}_{m}$ is the beam function. The matrix $\mathbf{Y}$ is defined so that it transforms the multipoles of the E- and B-modes into real-space maps of the Stokes parameters $Q$ and $U$. However, this method is computationally demanding, so we continue to use the simpler diagonal filtering approach in our work, as it provides a good approximation for our needs.

For the ground-based experiments, we will initially perform map combination using the inverse-variance method:
    \begin{equation}
		\begin{aligned}
			&\chi_{\ell m} = \sum_i \omega_{i, \ell} \chi_{i, \ell m}, \\
			&\omega_{i, \ell} = \frac{N_{i, \ell}^{-1}}{\sum_i N_{i, \ell}^{-1}} .
		\end{aligned}
	\end{equation}
 Here, $\omega_{i, \ell}$ represents the weights of the $i$-th experiment, $N_{i, \ell}$ is the E-mode noise power spectrum of the $i$-th experiment after beam deconvolution, $\chi_{i, \ell m}$ denotes the harmonic coefficients of fields $Q,U$ of the $i$-th experiment after beam deconvolution, and $\chi_{\ell m}$ represents the combined alm.

Next, we perform Wiener-filter on the combined maps, as mentioned above, and the combined E-mode noise power spectrum is given by:
	\begin{equation}
		N_{\ell}^{EE} = \sum_i \omega_{i, \ell}^2 N_{i, \ell}^{EE} .
	\end{equation}

\subsection{Lensing potential reconstruction}\label{phirec}

\begin{figure}
	\includegraphics[width=\columnwidth]{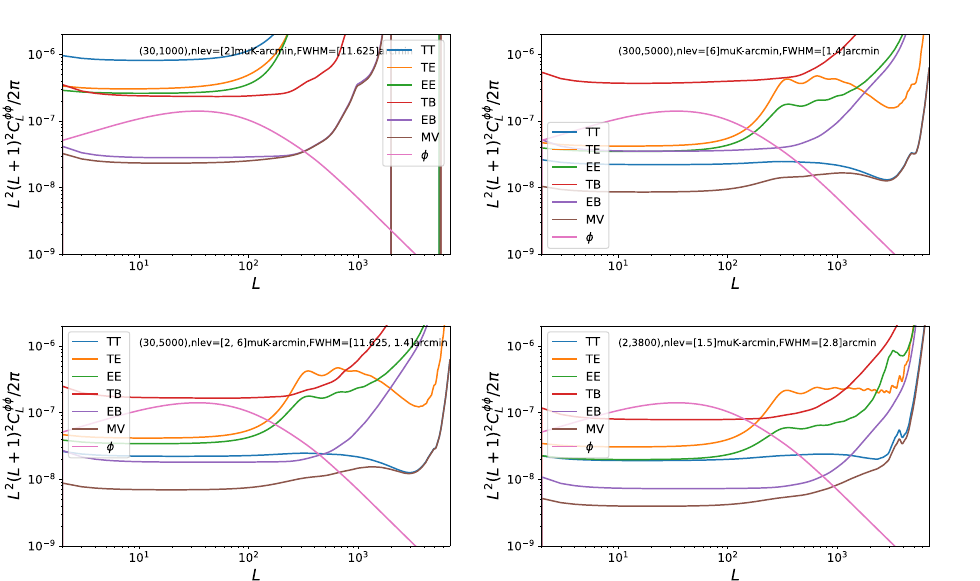}
	\caption{The lensing reconstruction noise power spectrum of the three experiments. The label of each curve represent the two CMB fields used to reconstruct. The noisiest $\phi$ is reconstructed from the gSAT (top-left), and combining it with the gLAT (top-right) leads to a reduction in reconstruction noise (bottom-left).
 The best performance is from the Satellite (bottom-right).}
	\label{fig:nlpp}
\end{figure}

\begin{figure}
	\includegraphics[width=\columnwidth]{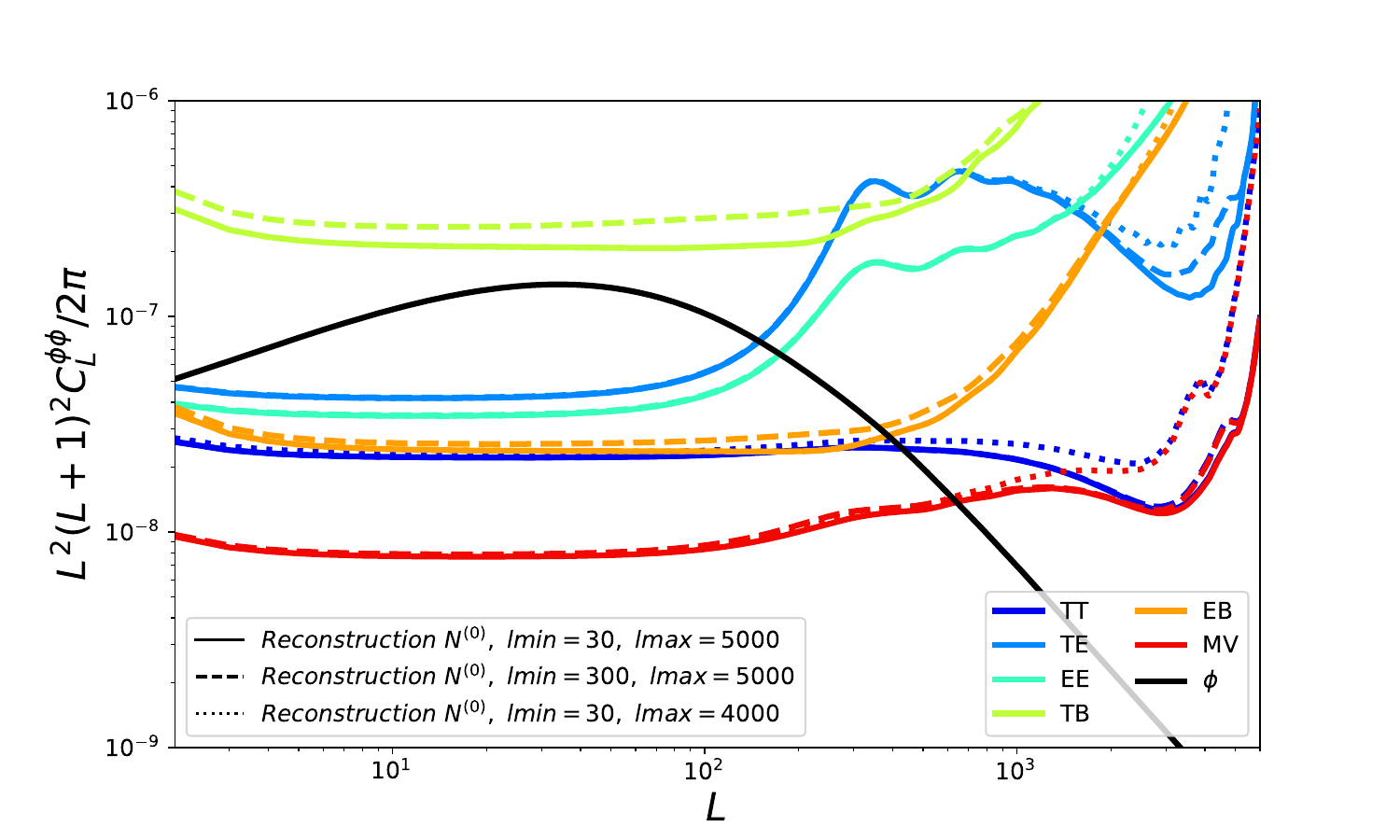}
	\caption{The lensing reconstruction noise power spectrum of the ground experiments with different multipole range of observation fields used for lensing reconstruction. Regarding MV estimators, compared to our baseline configuration (solid), excluding more large scale modes (dash) does not lead to an evident increase in $N^{(0)}$, while excluding more small scale modes (dot) does lead to an evident increase in $N^{(0)}$.}
	\label{fig:nlpp_lrange}
\end{figure}

In realistic situations, the lensing potential can be constructed either from internal datasets, such as the observed QU maps, or from external datasets like the large-scale structure \cite{smith2012delensing},\cite{sherwin2015delensing},\cite{manzotti2018future}. However, for the purposes of this discussion, we will only focus on the delensing process. Therefore, we will simply use the provided $\phi$ map with reconstruction noise, which is the Gaussian realization from the $N_l^{\phi\phi}$ given by the quadratic estimator:
\begin{equation}\label{EQ:nlpp}
	\begin{gathered}
		N^{(0),\alpha \alpha}_L = A^{\alpha}_L = (2L+1) \left\{ \sum_{l_1 l_2} \frac{|f_{l_1 L l_2}^{\alpha}|^2 }{2C_{l_1}^{aa}C_{l_2}^{bb}}\right\}^{-1}
	\end{gathered}
\end{equation}
where "$a$" and "$b$" denote one of the CMB field among {$\Theta$, E, B}, and $\alpha$ represents a pair of CMB field used for estimator, $f_{l_1 L l_2}^{\alpha}$ is the weight for the different quadratic pairs \cite{Okamoto:2003zw}, and $C_{l_1}^{aa}$ is the auto-power spectrum of a CMB field among {$\Theta$, E, B}. The noise spectra for different experiments are shown in Fig.\ref{fig:nlpp}. One can see that combining the gSAT with gLAT leads to a reduction in reconstruction noise thanks to the extension in small-scale coverage and the suppression of instrument noise. In realistic situation, large scale modes are usually excluded when reconstructing the quadratic estimator from E- and B-modes, to reduce the mode-overlap biases in the delensing procedure without losing too much signal-to-noise as the lensing reconstruction is mostly dominated by small scales (as shown in Fig.\ref{fig:nlpp_lrange}). Atmospheric noise and
Galactic foregrounds can also complicate the usability of gSAT data $(\ell< 300)$.  Additionally, extragalactic
foregrounds pose challenges for the gLAT when using temperature information \cite{osborne2014extragalactic}. These practical issues may degrade the performance of lensing reconstruction and the delensing process, which we plan to address in our future work.

Next, we perform Wiener-filtering on the $\phi$ maps:

\begin{equation}
	\begin{gathered}
		\phi_{LM}  \Rightarrow \frac{C_{\ell}^{\phi \phi}}{C_{\ell}^{\phi\phi} + N_{\ell}^{\phi\phi}} \phi_{LM}.
	\end{gathered}
\end{equation}

\subsection{Delensing}

Regarding the lensing template method, we proceed by converting the spin-2 fields $Q + iU$ to spin-1 and spin-3 fields, and transform the spin-0 lensing potential field to spin-$-1$ and spin-1 fields using \texttt{CMBlensplus}. It's important to note that only E-modes are included in the QU fields. Following this conversion, we compute the gradient lensing template according to Eq.(\ref{EQ17}) and Eq.(\ref{EQ18}). The resulting QU represents the lensing effect on E-mode, from which the lensing B-mode can be separated using harmonic transformation. We then directly subtract it from the observed B maps to obtain the delensed B maps. 

Regarding the Inverse-lensing method, we proceed by estimating the inverse deflection angle $ \beta $  from the filtered potential, following Eq.(\ref{EQ21}) using \texttt{CMBlensplus}. Subsequently, we remap the filtered observed QU maps to obtain the delensed QU maps. These are then subtracted from the observed QU maps to derive the noisy lensing template QU maps. The lensing template B map can be separated through harmonic transformation. 

We compute the angular power spectra of the delensed B-mode $C_{\ell}^{del}$ and of the instrumental noise $N_{\ell}^{BB}$ using the \texttt{NaMaster} code, and an apodized mask with edges trimmed is applied to mitigate edge effects caused by delensing. The delensing fraction (Eq.(\ref{eq:delenfrac1})) is calculated by the ratio of the difference between $C_{\ell}^{del}$ and $N_{\ell}^{BB}$ to the theoretical lensing B-mode power spectrum $C_{\ell}^{lens}$.
We also compute the cross-spectrum of the lensing B-mode template \(C_{\ell}^{LT \times \text{obs}}\) and its auto-power \(C_{\ell}^{LT}\) using the \texttt{NaMaster} code. The transfer function for \(C_{\ell}^{LT}\) is calculated as the ratio of the average power of the lensing B-mode template from signal-only simulation to the average lensed BB power. Besides, the transfer function for \(C_{\ell}^{LT \times \text{obs}}\) is similarly calculated by replacing the numerator with the cross-spectrum. We use these transfer functions to rescale \(C_{\ell}^{LT \times \text{obs}}\) and \(C_{\ell}^{LT}\) when fitting parameters.

As for the noise debias, we input the required CMB maps and the potential maps according to Section \ref{sec:forms_of_bias}, their power spectra are calculated using the \texttt{NaMaster} code.
We then compute the average of the bias power spectra obtained from 500 simulations to estimate the actual bias power spectra. These bias power spectra are then used appropriately in the L1, L2 and L3 models according to our specific needs as described in Section \ref{sec:posterior}.

\subsection{Results}\label{sec:main_result}

\begin{figure}
        \centering
        \subfigure{
	       \includegraphics[width=\columnwidth]{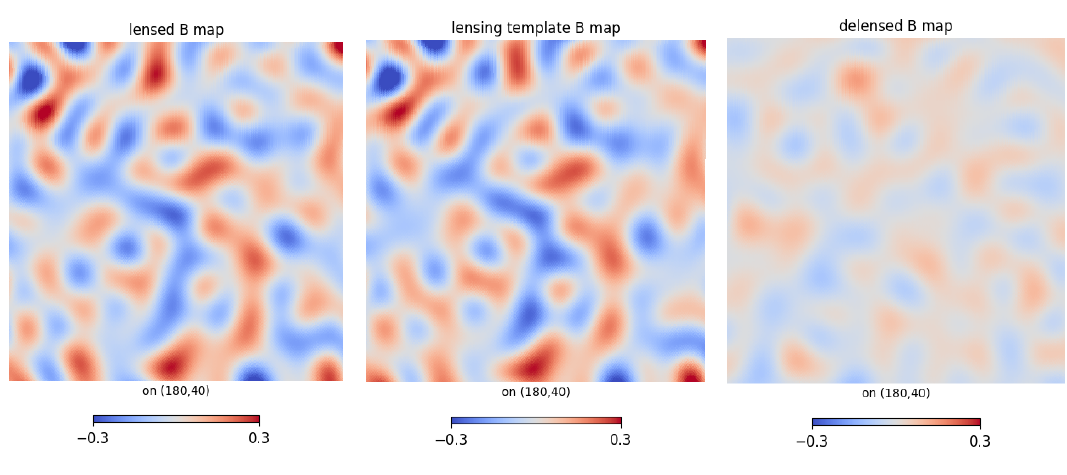}
	       \label{fig:temp_map}
        }
        \subfigure{
	       \includegraphics[width=\columnwidth]{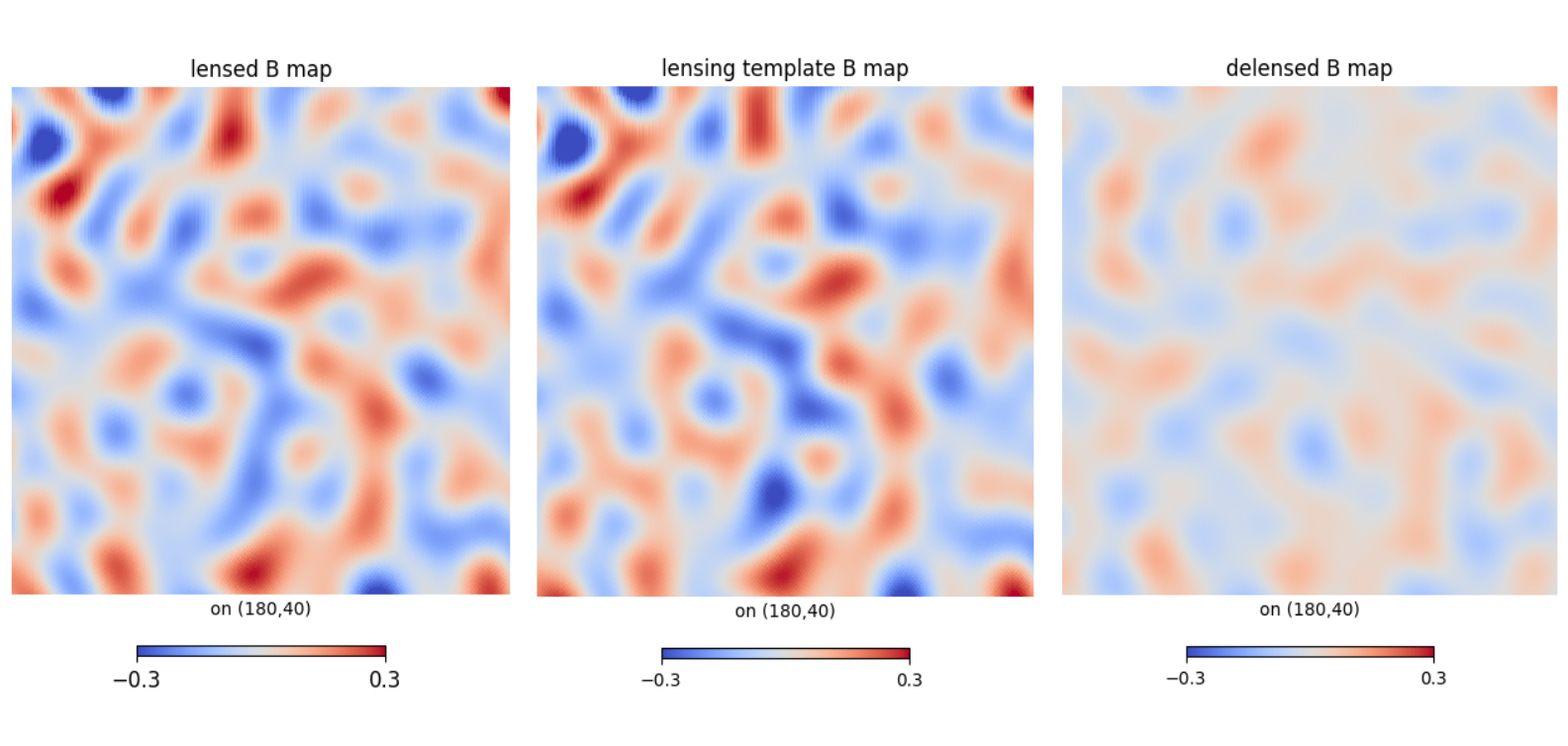}
	       \label{fig:remap_map}
        }
        \caption{The lensing template B map and delensed B map of the two delensing methods.
        The top three plots are from the gradient-order template method, and the bottom three are from the inverse-lensing method. We use the same input lensed B map for comparing the output. One can observe an obvious correlation between the input lensed B (left) and the lensing template B (medium), and the fading in the delensed B (right).}
\end{figure}

\begin{figure}
    \includegraphics[width=\columnwidth]{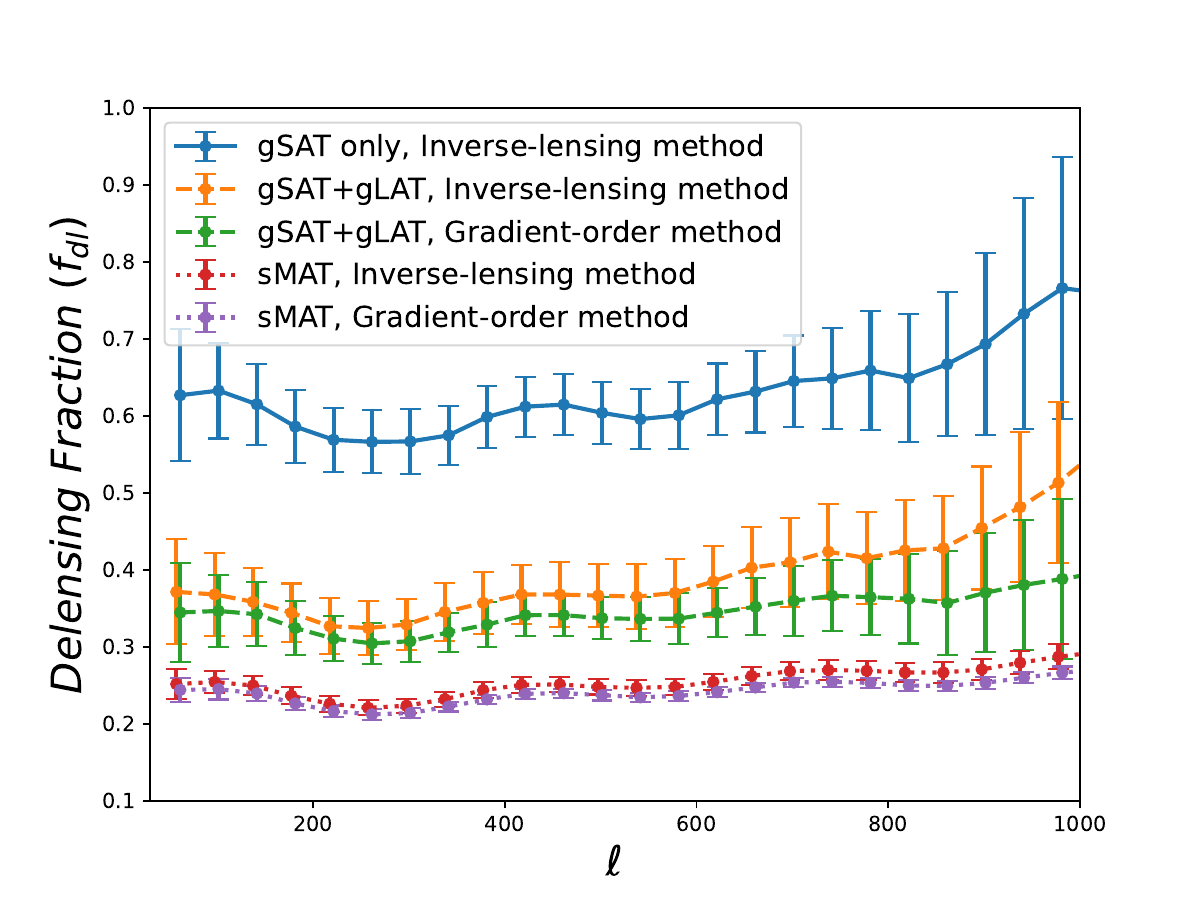}
	\caption{The delensing fraction of the three experiments with two methods. The solid blue shows gSAT delensing performance, 
	the orange and green dashed lines represent the delensing fraction with inverse-lensing method and gradient order template method if we combine gSAT and gLAT.
	We show the delensing performance of the sMAT with two method separately in red and purple dotted lines.}
	\label{fig:result}
\end{figure}

\begin{figure}
    \includegraphics[width=\columnwidth]{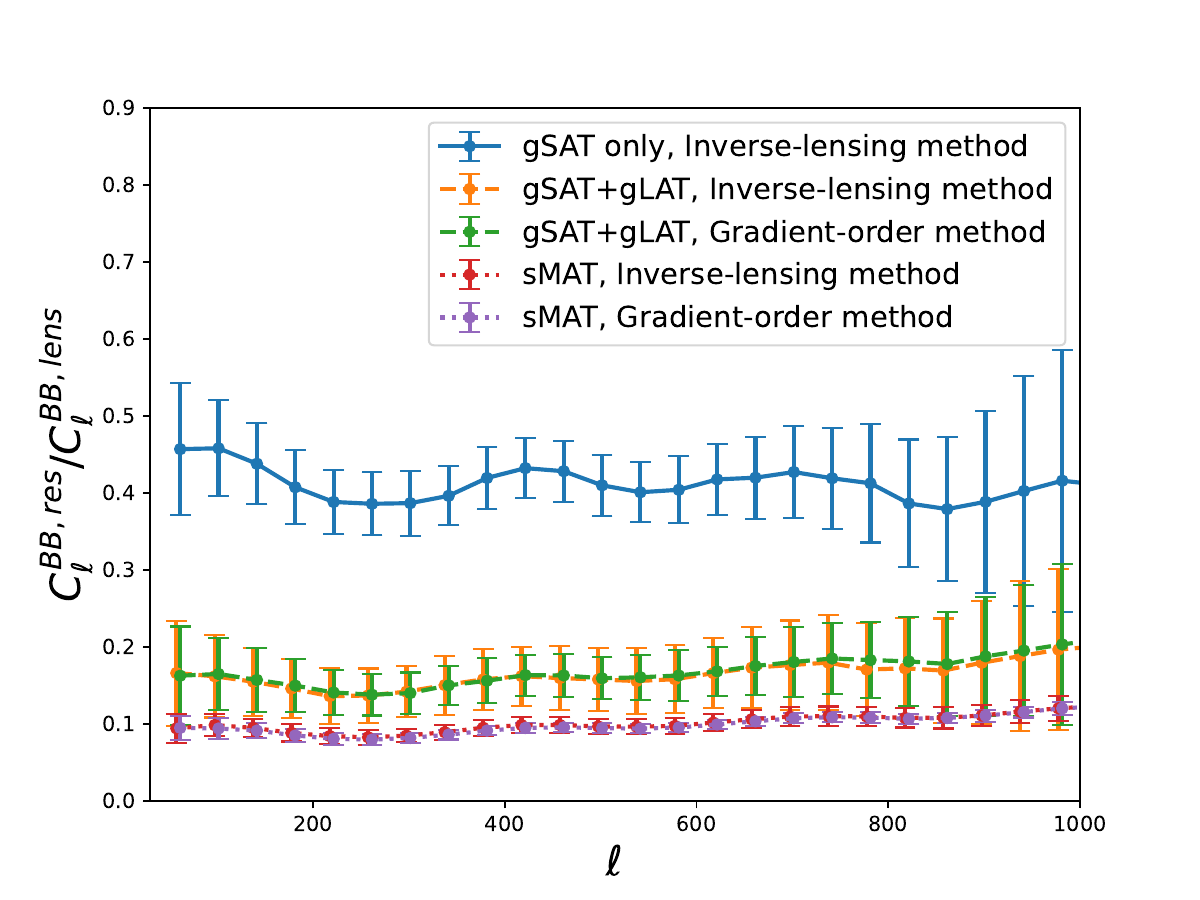}
	\caption{The ratio between lensing residual spectra and lensing spectra  of the three experiments with two methods. The solid blue is from gSAT, 
	the orange and green dashed lines are given by the inverse-lensing method and gradient order template method if we combine gSAT and gLAT, and the results of the sMAT with two method separately in red and purple dotted lines.}
	\label{fig:result_debiased}
\end{figure}

We plot the delensing fraction in Fig.\ref{fig:result}. We observe good consistency between the two methods for both experiments. 
However, the inverse-lensing method yields slightly higher values compared to the template method, this is due to the noise bias introduced during the delensing procedure, as discussed in \ref{sec:forms_of_bias}.
In Fig.\ref{fig:result_debiased}, where we artificially remove these high-order terms from the results, revealing perfect consistency between the two methods. This suggests that the delensing residuals introduced by the methods themselves are comparable. The primary difference between the two methods lies in noise-related power spectra introduced by the delensing procedure due to some high-order terms, which are mainly evident on small scales.

For the gradient method, utilizing Eq. (\ref{EQ2}), we derive $P^{temp} = \nabla P \cdot \nabla \phi$. Consequently (here we set $r=0$ for simplicity without loss of generality),

\begin{equation}\label{EQ:bias1}
    \begin{aligned}
              P^{del} &= P^{obs} - P^{temp} \\
                    &= [P^{len} + P^{noise}] - [\mathcal{W}^{E}\mathcal{W}^{\phi} \nabla(P^{len}+P^{noise})(\nabla\phi+\nabla\phi^{noise}) ] \\
                    &= \left\{ P^{len}-\mathcal{W}^{E}\mathcal{W}^{\phi} \nabla P^{len}\nabla\phi \right\}+ P^{noise} \\
                    &- \left\{\mathcal{W}^{E}\mathcal{W}^{\phi} \nabla P^{noise}(\nabla\phi+\nabla\phi^{noise})+\mathcal{W}^{E}\mathcal{W}^{\phi} \nabla P^{len}\nabla\phi^{noise}\right\},
    \end{aligned}
\end{equation}
where $P=Q \pm iU$  and this equation corresponds to Eq.(\ref{EQ26}), as the first brace corresponds to the residual B-mode, and the second brace corresponds to noise biases in Eq.(\ref{EQ26}).

As for the inverse-lensing method, from Eq.(\ref{EQ27}) and up to gradient order we have,
\begin{equation}\label{EQ:bias2}
    \begin{aligned}
        P^{del} &\approx \mathcal{W}^{E} (P^{len}+P^{noise})(n - \mathcal{W}^{\phi}(\nabla \hat \phi+\mathcal{O}(\nabla \hat \phi)^{2} )) \\
                &\approx \mathcal{W}^{E} (P^{len}+P^{noise})(n) \\
                &\quad- \mathcal{W}^{E}\mathcal{W}^{\phi} \nabla(P^{len}+P^{noise})(n) \nabla(\phi+\phi^{noise}) \\
                &= \left\{ \mathcal{W}^{E} P^{len}-\mathcal{W}^{E}\mathcal{W}^{\phi} \nabla P^{len}\nabla\phi \right\}+ \mathcal{W}^{E} P^{noise} \\
                &- \left\{\mathcal{W}^{E}\mathcal{W}^{\phi} \nabla P^{noise}(\nabla\phi+\nabla\phi^{noise})+\mathcal{W}^{{E}}\mathcal{W}^{\phi} \nabla P^{len}\nabla\phi^{noise}\right\},
    \end{aligned}
\end{equation}
Notice that we have neglected high-order terms since the second line, and $\mathcal{W}^{E} = C_{\ell}^{EE}/(C_{\ell}^{EE}+N_{\ell}^{EE})$ in the first term of the third line is close to unity because the E-mode signal is much stronger than its noise. Therefore, up to the gradient order, these delensing noise bias of the two methods is nearly equal. 
However, as discussed in Section \ref{sec:forms_of_bias}, it is the noise bias on small scales arising from the neglected high-order terms in Eq.(\ref{EQ:bias2}) that causes the delensing fraction obtained by the inverse-lensing method to slightly exceed that of the template method. We conclude that the inverse-lensing method is especially sensitive to noise, particularly at small scales.

The blue solid line indicates that using 0.8m-gSAT allows for the removal of approximately $40\%$ of the lensing B-modes power. This outcome is expected due to the large full width at half maximum (FWHM), which limits the observation of small-scale CMB polarization. Consequently, a significant reconstruction noise in the lensing potential occurs, degrading the delensing process.

Combining the 0.8m-gSAT with 6m-gLAT further enhances delensing by removing an additional $25\%$ of the lensing B-modes. This combination effectively reduces the FWHM, thereby decreasing the noise in the lensing potential as shown in Fig.\ref{fig:nlpp} and improving the signal-to-noise ratio in the observed CMB maps.

Future satellite experiments, as described in the paper, are able to provide a high signal-to-noise ratio on scales at $\ell < 3000$. This results in a high signal-to-noise ratio in the reconstruction of $\phi$, which is a major factor in the construction of the lensing template and the delensing procedure.
These experiments can remove approximately $80\%$ of the lensing B-modes using either of the two delensing methods.

Compared to Fig.\ref{fig:result}, the values of the delensing fractions
are greatly reduced by approximately 10\% to 20\% for the two methods in each experiment. This reduction is consistent with the statement made in Section \ref{sec:bias}, where we discussed that the delensing noise bias contributes to the final delensed B-modes and can be separated out from the power spectrum.
We observe a high degree of consistency in the fraction of the two methods, as mentioned above. This confirms the theoretical analysis indicating that the lensing template method and the inverse lensing method are almost equal in principle to some degree.

Furthermore, as discussed in Section \ref{sec:summary_of_bias}, achieving a smaller bias on $r$ requires flatter fractions. Notably, the lensing residual fractions exhibit a flatter behavior compared to the delensing fraction. Therefore, for future $r$ constraint with delensing, adopting the parameterization described in Eq.(\ref{eq:fitr2}) and utilizing our method to estimate the delensing noise spectra would be advisable. We will show the improvement in constraining 
$r$ in Section \ref{sec:posterior}.

\subsection{Comparison between the methods}


To date, several delensing methods have been proposed. In this section, we analyze and compare their characteristics based on our practical experience.

Firstly, we note from the detailed calculations presented in \cite{BaleatoLizancos:2020jby} that the gradient order template outperforms the non-perturbative order template, which is derived by remapping the lensed CMB fields with the lensing potential—a method not used in this work. This advantage arises from its ability to cancel out some high-order terms when computing the residual B-modes power spectrum. These higher-order terms indeed contribute to the delensing fraction, thereby leading to better performance of the gradient order template.


Furthermore, it is worth noting that the inverse-lensing method demonstrates comparable performance to gradient order template. The approximation $\beta = -\nabla \phi = -\alpha$ is widely used when computing the inverse deflection angle. Expanding on this approximation, we extend the analysis to include the second-order term.
\begin{equation}
	\begin{gathered}
		\beta(\hat n) = - \alpha(\hat n+\beta(\hat n)) = - \alpha(\hat n) - \nabla \alpha(\hat n) \cdot \beta(\hat n) - \dots.
	\end{gathered}
\end{equation}
We can see that neglecting the second-order term $\nabla \alpha(\hat n) \cdot \beta(\hat n)$ in $\beta$ results in retaining the second-order lensing effect in the remapped field. 
Consequently, the approximation $\beta = -\nabla \phi = -\alpha$ will lead to the persistence of high-order terms in the residual B-modes power spectrum, thereby  degrading its performance.

However, in our methodology, we employ an iterative approach instead of relying on the approximation. This iterative method effectively incorporates the second-order term $\nabla \alpha(\hat n) \cdot \beta(\hat n)$ into $\beta$. Consequently, it ensures that the second-order lensing effect is properly accounted for in the remapped field. As a result, the second terms in the residual B-modes power spectrum are canceled out, leading to comparable performance with the gradient order template method. For more details, please refer to \cite{BaleatoLizancos:2020jby}.

In terms of implementation efficiency, the non-perturbative template method stands out as the fastest. This is because we only need use \texttt{Lenspyx} to remap the lensed maps again with $\phi$ map, which is a relatively quick process.
The gradient order template method follows as the second fastest. While calculating the gradient of QU maps and potential maps isn't slow, 
the E-to-B leakage correction step \cite{liu2019methods} can be time-consuming when converting the lenisng QU template to lensing B template.
On the other hand, the inverse-lensing method is the slowest. Although calculating the inverse deflection angle $\beta(\hat n)$ isn't slow, remapping the lensed maps with $\beta(\hat n)$ using \texttt{CMBlensplus} proves to be memory- and time-consuming, leading to larger computational resource demand and longer processing times compared to the other two methods.

In conclusion, we recommend considering both the gradient order template and the inverse-lensing method for delensing tasks. 
The gradient order template is widely used, while the inverse-lensing method offers a more intuitive and straightforward conceptual framework.

However, it's important to note that while both methods can achieve comparable results on large scales, the inverse-lensing method typically requires more computational time due to the remapping process. 
Moreover, due to the definition of the gradient-order method, some parts of high-order effects cannot be captured by this method, particularly on small scales. Therefore, the inverse-lensing method should be more accurate on small scales compared to the gradient-order method. This will become increasingly evident as the future instrumental noise decreases, making its optimality evident. All in all, the choice of method should be systematic, considering not only the scales of interest but also the cost in terms of time and computational resources.

\subsection{Results on the $r$ constraint 
}\label{sec:posterior}

We investigate the parameter space using \texttt{Cobaya} \cite{Torrado_2021}, \cite{torrado2020cobaya}, which employs a Markov chain Monte Carlo approach. Three different Gaussian likelihoods are used to give parameter posterior distribution:

\begin{description}

    \item \textbf{L0:} Without applying the delensing process, just use the observed power spectrum $C_l$ directly with theoretical model given in Eq.(\ref{eq:fitr}).
    \item \textbf{L1:} The power spectrum $C_l$ described by the theoretical model given in Eq.(\ref{eq:fitr}).
    \item \textbf{L2:} The power spectrum $C_l$ described by the theoretical model given in Eq.(\ref{eq:fitr2}). 
    \item \textbf{L3:} Instead of applying the delensing process to the lensed maps, treat the lensing template as an additional pseudo-channel and calculate the auto and cross power spectra to create the theoretical model vector. Details can be found in Appendix \ref{app:l3}.
    
\end{description}

The Gaussian log-likelihood used in this work is given by:
\begin{equation}
	\begin{gathered}
        -2 \ \text{ln} \ \mathcal{L}(C_{\ell}|\hat{C_{\ell}}) = (\mathbf{X}_{\ell}-\mathbf{\hat X}_{\ell})^T \mathbf{M_{\ell}^{-1}} (\mathbf{X}_{\ell}-\mathbf{\hat X}_{\ell}) +  \text{ln} \ |\mathbf{M_{\ell}}|
	\end{gathered}
\end{equation}
where the vector $\mathbf{\hat X}_{\ell}$ contains the band-power estimates from the observed cut-sky maps, calculated using the \texttt{NaMaster} code. 
For ground-based and satellite experiments, we use the band-power spectrum from $\ell_{\text{min}} = 40$ to $\ell_{\text{max}} = 1000$ and from $\ell_{\text{min}} = 2$ to $\ell_{\text{max}} = 760$ respectively, with a binning width of $\Delta_\ell = 40$. This binning improves the Gaussianity of the distribution of $\hat C_{\ell}$, making the Gaussian approximation reliable according to \cite{hamimeche2008likelihood}.

The vector $\mathbf{X}_{\ell}$, which has components corresponding to $\mathbf{\hat X}_{\ell}$, represents the model calculations, with power spectra binned in the same way. The theoretical $C_{\ell}^{lens}$ and $N_{\ell}^{BB}$ used in our models are derived from the mean power spectrum of 500 sets of simulated cut-sky lensed B-maps and noise B-maps utilized in our baseline, allowing us to accurately capture any complexities.

More explicitly, $\mathbf{\hat X}_{\ell}$ corresponds to $C^{BB,obs}_{\ell}$ for case \textbf{L0} and to $C^{BB,del}_{\ell}$ for cases \textbf{L1} and \textbf{L2}. In case \textbf{L3}, where we include the Lensing B-mode template in the likelihood, $\mathbf{\hat{X}_{\ell}} = [C_{\ell}^{obs}, C_{\ell}^{LT \times obs}, C_{\ell}^{LT}]$. The details of corresponding models of $\mathbf{X}_{\ell}$ are given by Eq.(2.38), Eq.(2.39) and Appendix C, respectively.

Priors on $r$ and $A_L$ are imposed as uniform distributions $\mathcal{U}(-0.2,0.2)$ and $\mathcal{U}(0,1.2)$ respectively.

The covariance matrix $\mathbf{M_{\ell}}$ used in the likelihood is also derived from the baseline 500 simulations to account for the effects of masking, cosmic variance, noise, and other factors. Each simulation contains a CMB realization and noise realization, and we perform delensing on them to get the delensed B-maps and lensing B-map templates. We then calculate all power spectra contained in the aforesaid $\mathbf{\hat X}_{\ell}$ for each simulation using the \texttt{NaMaster} code as described in Section 4.3. 
The covariance matrix is then estimated from these 500 realizations of $\mathbf{\hat X}_{\ell}$, with the elements of the matrix:
\begin{equation}
    \mathbf{M}_{\ell,ij} = \langle \mathbf{\hat{X}}_{\ell,i} \mathbf{\hat{X}}_{\ell,j} \rangle - \langle \mathbf{\hat{X}}_{\ell,i} \rangle \langle \mathbf{\hat{X}}_{\ell,j} \rangle
\end{equation}
where $\mathbf{\hat{X}}_{\ell,i} = C_{\ell}^{obs}$ for case \textbf{L0}, $\mathbf{\hat{X}}_{\ell,i} = C_{\ell}^{del}$ for case \textbf{L1} and \textbf{L2}, and $\mathbf{\hat{X}}_{\ell,i} \in \{C_{\ell}^{obs}, C_{\ell}^{LT \times obs}, C_{\ell}^{LT}\}$ for case \textbf{L3}. The average is taken over 500 simulations.

\begin{figure}
 
    \includegraphics{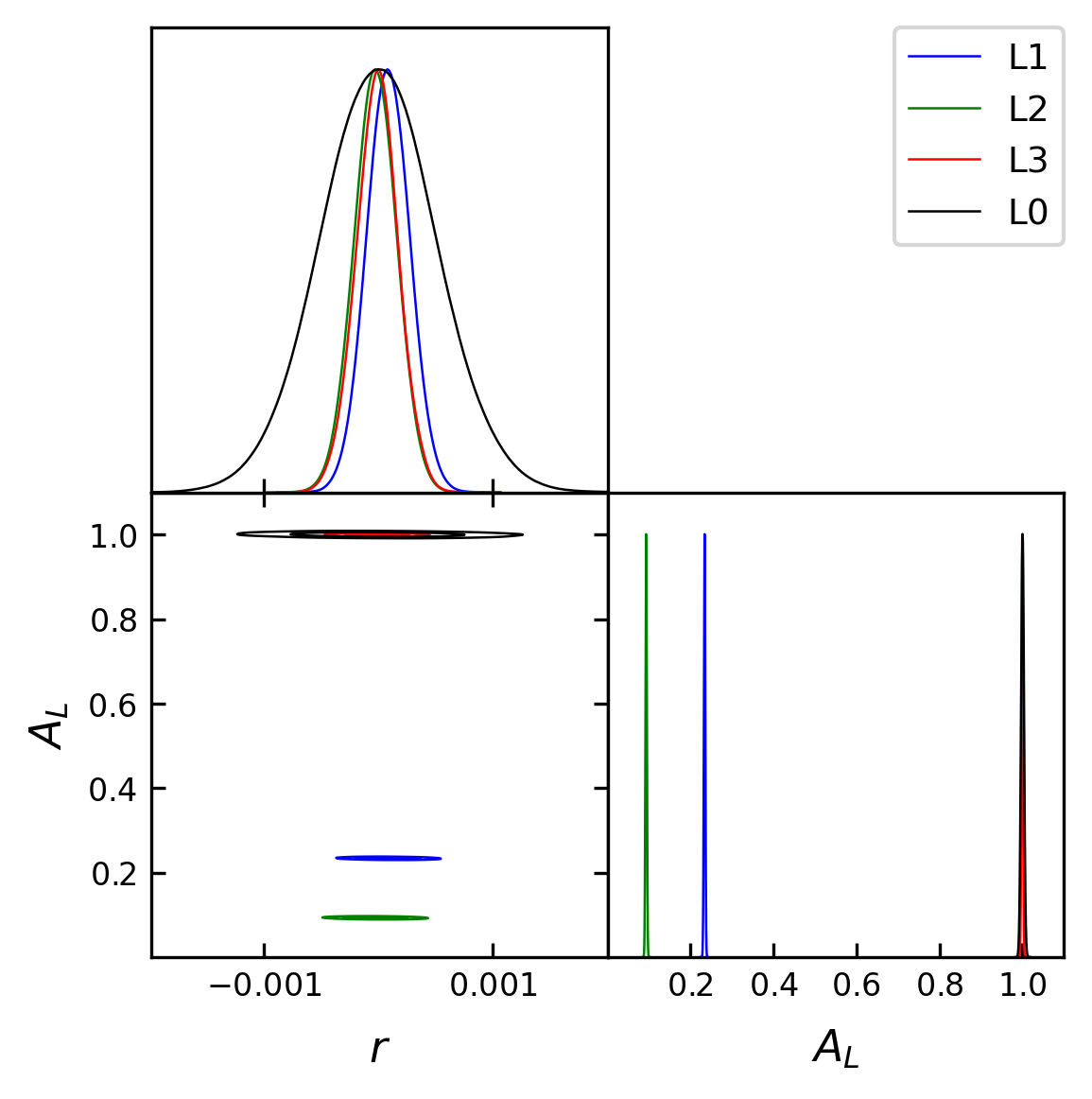}

    \includegraphics{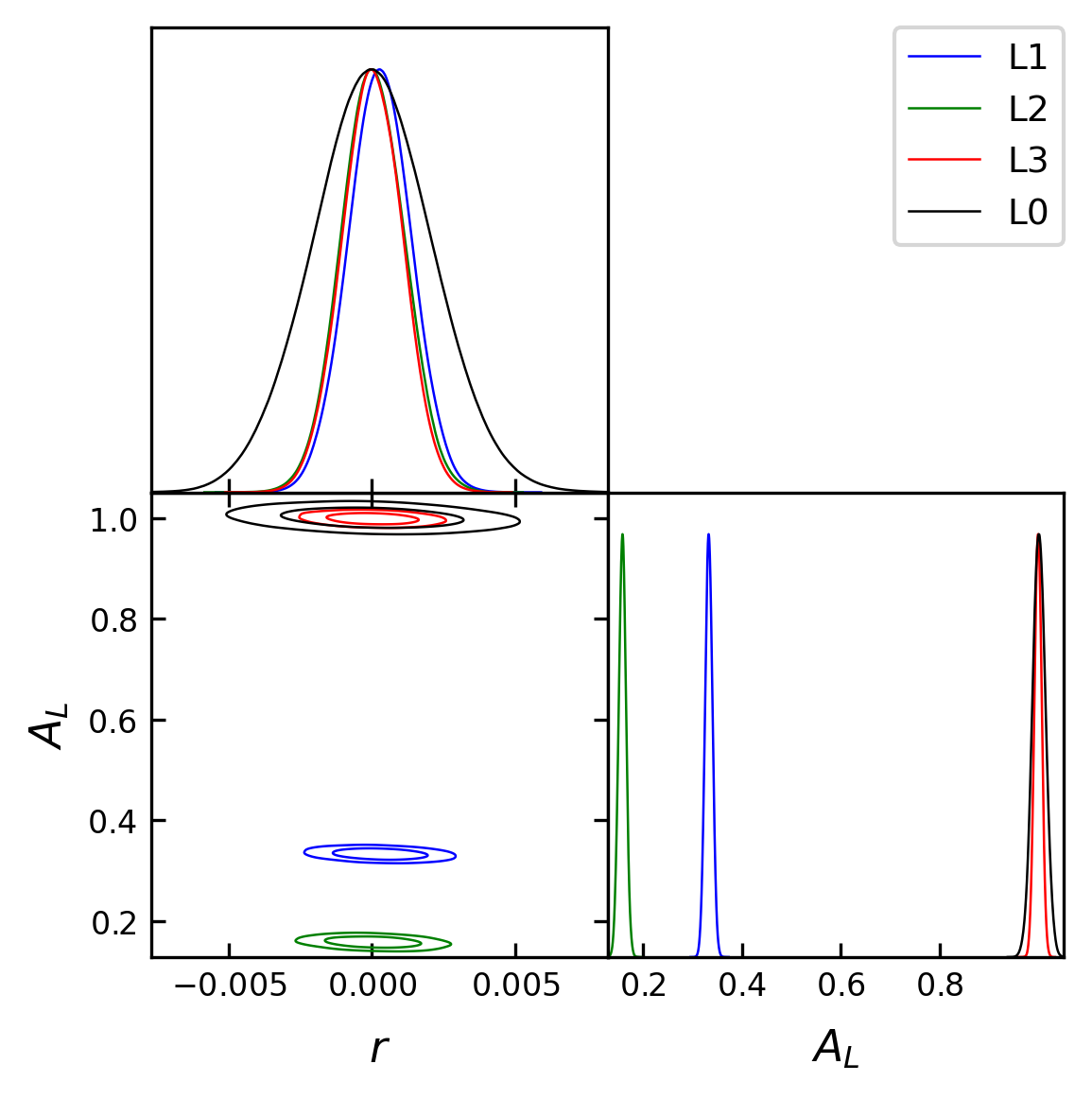}
    
	\caption{Posterior distributions on $r$ and $A_L$ given the baseline simulation data set. All three models with the akin $\sigma(r)$ reduction performance, while \textbf{L1} leads to the largest bias on $r$. With the addition of noise bias $N_{\ell}^{del}$ (\textbf{L2} in green), the peak of the distribution shifts towards $r=0$. Furthermore, adding LT as an independent channel leads to fewer bias on $r$ (\textbf{L3} in red). The upper plot is from the satellite experiment and the lower is from the gSAT+gLAT.}
	\label{fig:posterior}
\end{figure}

We present the posterior distributions for $r$ and $A_L$ using the 4 models with different data in Fig.\ref{fig:posterior}. The summarized results can be found in Table \ref{tab:posterior1} and Table \ref{tab:posterior2}. 
In each table, the first and second lines show the result using two models with delensed data (separately \textbf{L1} and \textbf{L2}), the third line shows the result adding the lensing template map as a new channel to analysis (\textbf{L3}). Additionally, we provide results based on the data without any delensing process for reference in the last line (\textbf{L0}). The parameter $A_L$ serves as an estimate of the delensing fraction; the results of $A_L$ in Table \ref{tab:posterior1} and Table \ref{tab:posterior2} align well with the outcomes depicted in Fig.\ref{fig:result} and Fig.\ref{fig:result_debiased} respectively.
For the satellite experiment, the delensing process leads to a reduction in $\sigma(r)$ from $0.504 \times 10^{-3}$ to $0.187 \times 10^{-3}$, corresponding to a reduction of approximately $\sim 63\%$. Similar reductions in $\sigma(r)$ due to delensing are observed in other experiments, with around a $46\%$ reduction in the gSAT+gLAT experiment.

Although map-based delensing process (\textbf{L1}) is beneficial in reducing $\sigma (r)$, it introduces bias on $r$. From Fig.\ref{fig:result}, we observe that the delensing fractions does not remain constant, while we incorporate this variability by fitting them with a floating constant parameter $A_L$ in our model. As we mentioned in Section \ref{sec:main_result}, we notice that the delensing fractions become more uniform in Fig.\ref{fig:result_debiased} after separating the delensing noise bias $N_{\ell}^{del}$, leading us to expect a reduction in the bias on $r$.
The result in Table \ref{tab:posterior2} also support that including $N_{\ell}^{del}$ in the model (\textbf{L2}) contributes to a reduction in the bias on $r$ compared with model (\textbf{L1}). 

One can see that from model \textbf{L2} to model \textbf{L1}, the bias on $r$ is significantly reduced, about 88\% for the ground experiments and 71\% for the satellite one.
Besides, the models \textbf{L1} and \textbf{L2} give the comparable $\sigma(r)$, which is expected since $\sigma(r)$ should be determined by the total delensed power spectrum, according to Eq.(\ref{eq:sigmar}). Fortunately, a more promising method (\textbf{L3}) is to add the lensing B-mode template as an independent channel to analysis, which leads to much lower bias on $r$ in both experiments with a compatible uncertainty reduction performance. This is because that the constructed lensing template is an unbiased estimate of the real lensing B-mode to gradient order, and the coefficient of $C_{\ell}^{lens}$ for each power spectra in \textbf{L3} is perfectly constant, the only bias comes from the noise spectra estimation which is much easier than the previous two models. 

Regarding the inverse lensing method, we find a similar performance in uncertainty reduction, though with slightly more bias. This is expected, as separating its bias terms requires a linear approximation. However, this should not have a significant impact compared to the remaining bias. We leave the investigation of a more precise separation of bias terms for future research.

\begin{table}
    \centering
    \caption{The mean and standard deviation of each parameter for gSAT+gLAT.}
    \label{tab:posterior1}
    \begin{tabular}{lcccc} 
        \hline
        \quad & \multicolumn{4}{c|}{gSAT+gLAT}  \\
        Model & r Mean ($10^{-3}$) & $\sigma(r) (10^{-3})$ & $A_L$ Mean & $\sigma({A_L})$ \\
        \hline
        \textbf{L1} & 0.276 & 1.104 & 0.333 & 0.007 \\
        \textbf{L2} & 0.033 & 1.110 & 0.159 & 0.007 \\
        \textbf{L3} & 0.011 & 1.055 & 0.999 & 0.007\\
        \textbf{L0} & 0.003 & 2.073 & 1.000 & 0.013\\
        \hline
    \end{tabular}
\end{table}

\begin{table}
    \centering
    \caption{The mean and standard deviation of each parameter for sMAT.}
    \label{tab:posterior2}
    \begin{tabular}{lcccc} 
        \hline
        \quad & \multicolumn{4}{c|}{sMAT}  \\
        Model & r Mean ($10^{-3}$) & $\sigma(r) (10^{-3})$ & $A_L$ Mean & $\sigma({A_L})$ \\
        \hline
        \textbf{L1} & 0.087 & 0.188 & 0.233 & 0.002 \\
        \textbf{L2} & -0.025 & 0.187 & 0.093 & 0.002\\
        \textbf{L3} & -0.008 & 0.185 & 1.000 & 0.002 \\
        \textbf{L0} & 0.001 & 0.504 & 1.000 & 0.004\\
        \hline
    \end{tabular}
\end{table}

\subsection{Discussion on the noise of the $\phi$ map}\label{sec:phi_noise}
Here we analyze the impact of utilizing a Gaussian $\phi$ noise, as discussed in Section \ref{phirec}. Firstly, the equation is derived under the assumption that the lensed CMB is a Gaussian random field on the sphere, considering it as the primary source of error in the lensing reconstruction. However, higher-order noise biases $N_L^{(p)}$ can also significantly affect the results, as highlighted in \cite{hanson2011cmb}. This implies that we may be underestimating the reconstruction noise bias compared to practical lensing reconstruction, particularly at $L>1000$ where $N_L^{(1)}$ dominates the reconstruction power spectrum.
Secondly, Gaussian realizations of the dominant reconstruction error $N_L^{(0)}$ are generated and then added to a Gaussian $\phi$ map. It is important to note that in practical applications, the quadratic estimator $\hat{\phi}$ is non-Gaussian. The introduction of an internal $\phi$ proxy has the potential to adversely impact delensing performance due to the presence of additional non-Gaussian correlation terms. In our current analysis, this Gaussian $\phi$ noise realization does not correlate with the lensed CMB, so we neglect the correlated terms on the delensing process. Future studies will delve deeper into the intricacies associated with the use of an internal $\phi$ tracer, where further considerations will be made to address the potential effects of non-Gaussianity and correlated terms on the delensing process. 

\section{Discussion and Conclusion}\label{sec:conclusion}
In this paper, we demonstrate CMB B-mode delensing through gradient-order and inverse-lensing methods, offering a comprehensive analysis of both the delensing residual originating from the methods themselves and the delensing noise biases stemming from instrumental noise and those from the reconstruction of $\phi$.
We investigate the delensing fraction using simulated data obtained from both small-aperture and larger-aperture ground-based telescopes, as well as space missions equipped with medium-aperture instruments. 
In addition, we provide constraints on $r$ with different likelihood functions, illustrating the impact of delensing in reducing its uncertainty. 

The results reveal that a ground-based small-aperture CMB polarizing telescope achieves a stand-alone lensing B-mode removal efficiency of 40\%. This efficiency increases to 65\% when combined with a large-aperture telescope. Furthermore, the future satellite experiment achieves an impressive removal efficiency of approximately 80\%. Both delensing methods can attain comparable lensing residuals. 
However, even though the inverse-lensing method is more accurate on small scales, it typically demands more computational time due to its remapping process and shows no significant discrepancy compared to the more economical gradient-order method on large scales at the current level of instrumental noise. Even so, we still believe the inverse-lensing method is the optimal one, and its advantages will become more apparent as the instrumental noise levels decrease in future experiments.

Four different likelihood functions are used to constrain the $r$ parameter. The baseline likelihood (\textbf{L0}), without any delensing process, provides an almost unbiased estimate of 
 $r$ but with the largest uncertainty. The likelihood (\textbf{L1}), where we parameterize the delensing fraction ($f_{dl}$) as $A_L$, results in a reduction of $\sigma(r)$ by approximately 46\% for ground-based experiments. This reduction further improves to 63\% for space missions. However, this approach introduces a noticeable bias in $r$, even though it remains within $1 \sigma$. We introduce an alternative approach (\textbf{L2}), where we parameterize the fraction ($C_l^{res}/C_l^{lens}$) as $A_L$ instead of $f_{dl}$ and estimate $N_l^{del}$ based on simulations. This method reduces biases by 88\% for the gSAT+gLAT experiment and by 71\% for the satellite experiment, with almost the same $\sigma(r)$. A more promising method (\textbf{L3}) is to add the lensing template B-mode as an additional frequency channel, which leads to negligible bias on $r$ in both experiments with a compatible uncertainty reduction performance. 
Above all, we believe the promising and economical way to improve the sensitivity to $r$ at current instrument condition is the combination of gradient-order method and \textbf{L3} likelihood.

We have noticed that the galactic and extragalactic foregrounds will bias the delensing procedure, which is also widely discussed by \cite{beck2020impact}. We plan to further examine the influence of the foregrounds on our methods in a future paper.


 We will enhance our delensing pipeline by integrating internal lensing potential reconstruction techniques in the future work. Additionally, we plan to combine the lensing potential maps with other $\phi$ tracers, such as the Cosmic Infrared Background and Galaxy number density \cite{sherwin2015delensing}, \cite{manzotti2018future}, \cite{namikawa2024litebird} to increase the signal-to-noise ratio of the lensing potential and improve the delensing efficiency.
What's more, we notice that recently the iterative internal delensing method has been performed in \cite{belkner2023cmb}, and we will further compare the performance of the iterative delensing method with the two methods in the future work.
Additionally, we will provide a more comprehensive analysis of delensing performance across the three experiments, accounting for foreground emissions and inhomogeneous noise.

\appendix
\section{Some algorithms of delensing effect}\label{sec:algorithms}
Here we want to highlight an algorithm of delensing effect, which may be useful when separating the components of the delensing output. 
We first assume that we remap a lensed field with a potential field $\phi = \phi_1+\phi_2$ which is artificially divided into two components (e.g. signal and noise), corresponding to inverse-lensing angle $\beta_1$ and $\beta_2$, respectively.
We assume the total inverse-lensing angle $\beta \approx  \beta_1 + \beta_2$, we have to admit that this neglects the second order term of $\beta$ (the gradient operation is linear, while the iteration is not), 
but it seems not a big deal to our final results.
Then we can write the delensed field as:
\begin{equation}
	\begin{aligned}
		X^{de}(\hat n) &\approx  \tilde X (\hat n + \beta_1 + \beta_2) \\
		& \approx \tilde X(\hat n ) + \nabla \tilde X (\hat n) ( \beta_1 + \beta_2) + \frac{1}{2} \nabla^2 \tilde X (\hat n) (\beta_1 + \beta_2)^2 + \mathcal{O} (\beta ^2)\\
		& \approx \left[\tilde X(\hat n)  +  \nabla \tilde X(\hat n) \beta_1 +  \frac{1}{2} \nabla^2 \tilde X(\hat n) \beta_1^2\right] \\ 
			&+ \left[ \tilde X(\hat n) + \nabla \tilde X(\hat n) \beta_2  + \frac{1}{2} \nabla^2 \tilde X(\hat n) \beta_2^2\right] - \tilde X(\hat n )+ \mathcal{O} (\beta ^2)\\
		& \approx \tilde X(\hat n + \beta_1) + \tilde X(\hat n + \beta_2) - \tilde X(\hat n ) + \mathcal{O} (\beta ^2).
	\end{aligned}
\end{equation}

However, if we remap a sum of two lensed field $\tilde X = \tilde X_1 + \tilde X_2$ with a single potential, it will be easy finding that this just a linear algorithm, so:
\begin{equation}
	\begin{aligned}
		X^{de}(\hat n) = \tilde X(\hat n+\beta) =  X_1(\hat n + \beta) +  X_2(\hat n + \beta) = X_1^{de}(\hat n ) +  X_2^{de}(\hat n ).
	\end{aligned}
\end{equation}

Above all, we can summary here the delensing algorithms :
\begin{equation}\label{EQA3}
	\begin{aligned}
		&X \star \left(\phi_1 + \phi_2\right) \approx  X \star \phi_1 + X \star \phi_2 - X + \mathcal{O} (\beta ^2), \\
		&(X_1 + X_2) \star \phi =  X_1 \star \phi + X_2 \star \phi.
	\end{aligned}
\end{equation}
Here, we use $\star$ to represent the delensing operation, $X$ represents one of  $\{ \Theta, Q \pm iU \}$ among the CMB field, instrumental noise, or the observed CMB field, and $\phi$ represents the lensing potential or its reconstruction noise.

As for the gradient order template method, we find the lensing template map is straightforward and just obeys a linear algorithm as :
\begin{equation}
	\begin{aligned}
		&\nabla[(Q_1+Q_2)\pm i(U_1+U_2)](\hat{n})\nabla[\phi_1(\hat{n})+\phi_2(\hat{n})] \\
		&= \nabla[Q_1\pm iU_1](\hat{n})\nabla[\phi_1(\hat{n})] + \nabla[Q_2\pm iU_2](\hat{n})\nabla[\phi_2(\hat{n})] \\
		& \ + \nabla[Q_2\pm iU_2](\hat{n})\nabla[\phi_1(\hat{n})] + \nabla[Q_1\pm iU_1](\hat{n})\nabla[\phi_2(\hat{n})].
    \end{aligned}
\end{equation}

\section{Wiener filter}\label{sec:filter}
\subsection{Definition and derivation} 
\quad The well-known form of Wiener filter is written as: 
\begin{equation}
	\frac{S}{S+N},
\end{equation}
where $S$ represents the power spectrum of signal, and $N$ is the power spectrum of noise.

Here we will derive it from the delensing aspect. We provide two kind of understanding.
We will see that the principle of Wiener filter is to minimize the residuals (or called error).
\subsubsection{Derivation using gradient template method}
\quad We work in flat sky for simplicity. We start from the observed B-modes, which consists of PGWs induced B-modes, lensing induced B-modes and noise
\begin{equation}
	B^{obs}(l) = B^{tens}(l) + B^{lens}(l) + B^{noise}(l).
\end{equation}

Learning that the lensed B-modes can be written as follows (up to gradient order)
\begin{equation}
	B^{lens}(l) = -\int \frac{d^2l'}{2\pi} sin2(\psi_{l'}-\psi_l) l' \cdot (l-l') E(l') \phi(l-l').
\end{equation}

We write the lensing template B-modes by replacing E-modes and $\phi$ above with weighted observed E-modes and $\phi$ respectively as
\begin{equation}
    \begin{aligned}
        	B^{temp}(l) = -\int \frac{d^2l'}{2\pi} &sin2(\psi_{l'}-\psi_l) l' \cdot (l-l') \\
        &\times \mathcal{W}_{l'}^E E^{obs}(l') \mathcal{W}_{|l-l'|}^{\phi} \phi^{obs}(l-l').
    \end{aligned}
\end{equation}

Then the delensed B-modes can be written as
\begin{equation}
	\begin{gathered}
		B^{del}(l) = B^{obs}(l) - B^{temp}(l) = B^{tens}(l) + B^{noise}(l) + B^{res}(l),
	\end{gathered}
\end{equation}
where the residual B-modes is defined as:
\begin{equation}
	\begin{gathered}
		B^{res}(l) = B^{lens}(l) - B^{temp}(l) = -\int \frac{d^2l'}{2\pi} sin2(\psi_{l'}-\psi_l) l' \cdot (l-l') \\ \times \left[ E(l') \phi(l-l') - \mathcal{W}_{l'}^E E^{obs}(l') \mathcal{W}_{|l-l'|}^{\phi} \phi^{obs}(l-l') \right].
	\end{gathered}
\end{equation}

Then we can write the power spectrum of residual B-modes as
\begin{equation}
	\begin{aligned}
		C_l^{res} \cdot (2\pi)^2 &= \langle B^{res}(l) B^{res*}(l) \rangle \\
		&= \int \frac{d^2l_1}{2\pi} \int \frac{d^2l_2}{2\pi} \sin 2(\psi_{l_1}-\psi_l) \sin 2(\psi_{l_2}-\psi_l) \\
		&\times \left[ l_1 \cdot (l-l_1)\right] \left[ l_2 \cdot (l-l_2)\right] \\
		&\times \langle \left[ E(l_1) \phi(l-l_1) - \mathcal{W}_{l_1}^E E^{obs}(l_1) \mathcal{W}_{|l-l_1|}^{\phi} \phi^{obs}(l-l_1) \right] \\
		\times &\left[ E^*(l_2) \phi^*(l-l_2) - \mathcal{W}_{l_2}^E E^{obs*}(l_2) \mathcal{W}_{|l-l_2|}^{\phi} \phi^{obs*}(l-l_2) \right] \rangle.
	\end{aligned}				
\end{equation}

Use Wick's theorem and notice that 
\begin{equation}
	\begin{aligned}
		&\langle E(l_1)E^*(l_2) \rangle = (2\pi)^2 \delta^2(l_1-l_2) C_{l_1}^{EE},\\
		&\langle \phi(L_1)\phi^*(L_2) \rangle = (2\pi)^2 \delta^2(L_1-L_2) C_{l_1}^{\phi \phi },\\
		&\langle E(l_1)n^{E*}(l_2) \rangle = 0 \Rightarrow \langle E(l_1)E^{obs*}(l_2) \rangle = \langle E(l_1)E^*(l_2) \rangle,\\
		&\langle \phi(L_1)n^{\phi*}(L_2) \rangle = 0 \Rightarrow \langle \phi(L_1)\phi^{obs*}(L_2) \rangle = \langle \phi(L_1)\phi^*(L_2) \rangle, \\
		&\langle E^{obs}(l_1)E^{obs*}(l_2) \rangle = (2\pi)^2 \delta^2(l_1-l_2) \left[C_{l_1}^{EE} + N_{l_1}^{EE} \right],\\
		&\langle \phi^{obs}(L_1)\phi^{obs*}(L_2) \rangle = (2\pi)^2 \delta^2(L_1-L_2) \left[C_{L_1}^{\phi \phi} + N_{L_1}^{\phi \phi} \right],
	\end{aligned}				
\end{equation}

Substitute the above equations into the equation of $C_l^{res}$, and questing for the minimum of $C_l^{res}$, we can get the Wiener filter as
\begin{equation}
	\begin{aligned}
		\frac{\partial C_l^{res}}{\partial \mathcal{W}_{l_1}^{E}\mathcal{W}_{|l-l_1|}^{\phi } } &= \int \frac{d^2l_1}{2\pi} g^2(l_1,l) [-C_{l_1}^{EE}C_{|l-l_1|}^{\phi \phi}  \\
		&+ \mathcal{W}_{l_1}^{E}\mathcal{W}_{|l-l_1|}^{\phi} (C_{l_1}^{EE}+ N_{l_1}^{EE})(C_{|l-l_1|}^{\phi \phi}+N_{|l-l_1|}^{\phi \phi}) ] = 0,\\
		&\Rightarrow \mathcal{W}_{l_1}^{E}\mathcal{W}_{|l-l_1|}^{\phi} = \frac{C_{l_1}^{EE}}{C_{l_1}^{EE}+N_{l_1}^{EE}} \cdot \frac{C_{|l-l_1|}^{\phi \phi}}{C_{|l-l_1|}^{\phi \phi}+N_{|l-l_1|}^{\phi \phi}}.
	\end{aligned}				
\end{equation}

Therefore, the lensing template we construct is
\begin{equation}
	\begin{aligned}
		B^{temp}(l) &= -\int \frac{d^2l'}{2\pi} sin2(\psi_{l'}-\psi_l) l' \cdot (l-l') \mathcal{W}_{l'}^E E^{obs}(l') \\ &\times \mathcal{W}_{|l-l'|}^{\phi} \phi^{obs}(l-l')\\
			&= -\int \frac{d^2l'}{2\pi} sin2(\psi_{l'}-\psi_l) l' \cdot (l-l') \\ &\times \left[\frac{C_{l_1}^{EE}}{C_{l_1}^{EE}+N_{l_1}^{EE}} E^{obs}(l') \right] \left[\frac{C_{|l-l_1|}^{\phi \phi}}{C_{|l-l_1|}^{\phi \phi}+N_{|l-l_1|}^{\phi \phi}} \phi^{obs}(l-l')\right].
	\end{aligned}
\end{equation}

Obviously, this filtered template is a scaled version of the lensed B-modes due to the suppression of Wiener filter on signal.

Notice that if the beam effect is not contained in the noise power spectrum, the Wiener filter should be modified as follows:

\begin{equation}
	\begin{aligned}
		&E^{obs}_{\ell m } = E_{\ell m }b_{\ell} + n^{E}_{\ell m }, \\
		&\Rightarrow 
		\mathcal{W}^E_{\ell} E^{obs}_{\ell m } = b_{\ell}\mathcal{W}^E_{\ell} \left( E_{\ell m } + n^{E}_{\ell m }/b_{\ell} \right), \\
		&\Rightarrow 
		b_{\ell} \mathcal{W}^E_{\ell} = \frac{C_{\ell}^{EE}}{C_{\ell}^{EE}+N_{\ell}^{EE}/b_{\ell}^2} \Rightarrow  \mathcal{W}^E_{\ell} = \frac{1}{b_{\ell}} \frac{C_{\ell}^{EE}}{C_{\ell}^{EE}+N_{\ell}^{EE}/b_{\ell}^2},
	\end{aligned}
\end{equation}
where we just replace $b_{\ell} \mathcal{W}^E_{\ell}$ with original $ \mathcal{W}^E_{\ell}$, and the beam effect is not contained in the noise power spectrum.

\subsubsection{Derivation using inverse-lensing method}
\quad We try to follow \cite{Green:2016cjr} to derive the Wiener filter from another perspective.  This article is recommended for readers interested in gaining a more thorough understanding of the Wiener filter used in delensing procedure.
Here we only give a brief introduction, more detailed calculation can be find there.

For an observed field, different multipoles may have different signal-to-noise ratio, while most of the parts are signal dominated, some part can be corrupted by noise.
So, we artificially divide the observed field into two parts, one is signal dominated, the other is noise dominated, with three filter operators $h$, $\bar h$ and $g$.
We want the signal-dominated part to be remapped to the unlensed CMB, while the noise-dominated part remains unchanged to avoid any contamination of the result.
Then the delensed field can be written as (we use temperature field for instance, and can be generalized to QU fields without losing generality) 

\begin{equation}
	\begin{gathered}
		T^d(\hat x) = \bar h \star T^{obs}(\hat x) + h \star T^{obs}(\hat x - g \star \alpha^{obs}(\hat x)).
	\end{gathered}
\end{equation}
Where $\star$ represents the convolution. As for the reason of division, we leave it to the next subsection.
This is actually the inverse-lenisng method we introduced before, and we only remap the second part of the observed field which are signal-dominated, 
and leave the noise-dominated part unchanged artificially. We expect this choice will improve the delensing by improving the SNR of the input data.

We know that lenisng conserves the total power, so we here obey this rule to mimic the lensing effect since inverse-lensing is essentially also a kind of lensing
\begin{equation}
	\begin{gathered}
		\langle T^d(0)^2 \rangle = \langle T^{obs}(0)^2 \rangle .
	\end{gathered}
\end{equation}
After a short calculation, we can get the relation between $h$ and $\bar h$
\begin{equation}
	\begin{gathered}
		\bar h_{\ell} = \sqrt{1-h_{\ell}^2 \left( 1- e^{-\frac{\ell^2}{2} C_0^{obs}(0)} \right)} - h_{\ell}e^{-\frac{\ell^2}{4} C_0^{obs}(0)}.
	\end{gathered}
\end{equation}
Where $C_0(r) = \langle \alpha(\hat x) \cdot \alpha(\hat x') \rangle$ is the correlation function of deflection angle.
The filter are derived demanding the minimize the variance below (B.7) of \cite{Green:2016cjr}
\begin{equation}
	\begin{gathered}
		\langle (T_{\ell}^d - \langle T_{\ell}^d \rangle_{\phi,\phi^N})(T_{-\ell}^d - \langle T_{-\ell}^d \rangle_{\phi,\phi^N}) \rangle_{T,\phi,\phi^N}.
	\end{gathered}
\end{equation}
\quad An intuition for this choice is that we are demanding that we minimize how much $T^d$  varies with each realization of the lensing field and the reconstruction noise.
More explicitly, this actually corresponds to demanding the minimum of residual power, which is defined as the difference between the lensed field and the template field.
Once we get the delensed field, the template field will just be the difference between the lensed field and the delensed field.
\begin{equation}
	\begin{gathered}
		T^{del}(l) = T^{obs}(l) - T^{temp}(l) = T^{tens}(l) + T^{noise}(l) + T^{res}(l),
	\end{gathered}
\end{equation}
If we first average over the lensing potential and its noise, then $T^{tens}(l)$ and $T^{noise}(l)$ will not change and $(T_{\ell}^d - \langle T_{\ell}^d \rangle_{\phi,\phi^N})$ 
thus equal to $(T_{\ell}^{res} - \langle T_{\ell}^{res} \rangle_{\phi,\phi^N})$, then we average its norm square over all field realizations, we can get the variance of the residual field.
So, here we are actually demanding the minimum of residuals also, but in a more indirect way.

After some complicated calculation with explicit expression, we get the final output as 
\begin{equation}
	\begin{gathered}
		g_L = \frac {C_L^{\phi \phi}}{C_L^{\phi \phi} + N_L^{\phi \phi}}, \\
		h_{\ell} = \frac{\tilde C_{\ell}^{T\nabla T}}{C_{\ell}^{obs}} \simeq \frac{\tilde C_{\ell}^{T T}}{C_{\ell}^{obs}}.
	\end{gathered}
\end{equation}

\subsection{Why do we need Wiener filter in CMB delensing} 
\quad A brief and mainstream answer is to minimize the residuals by improving the input SNR.
Here we give a more detailed explanation.

\begin{figure}
	\centering
	\includegraphics[width=\columnwidth]{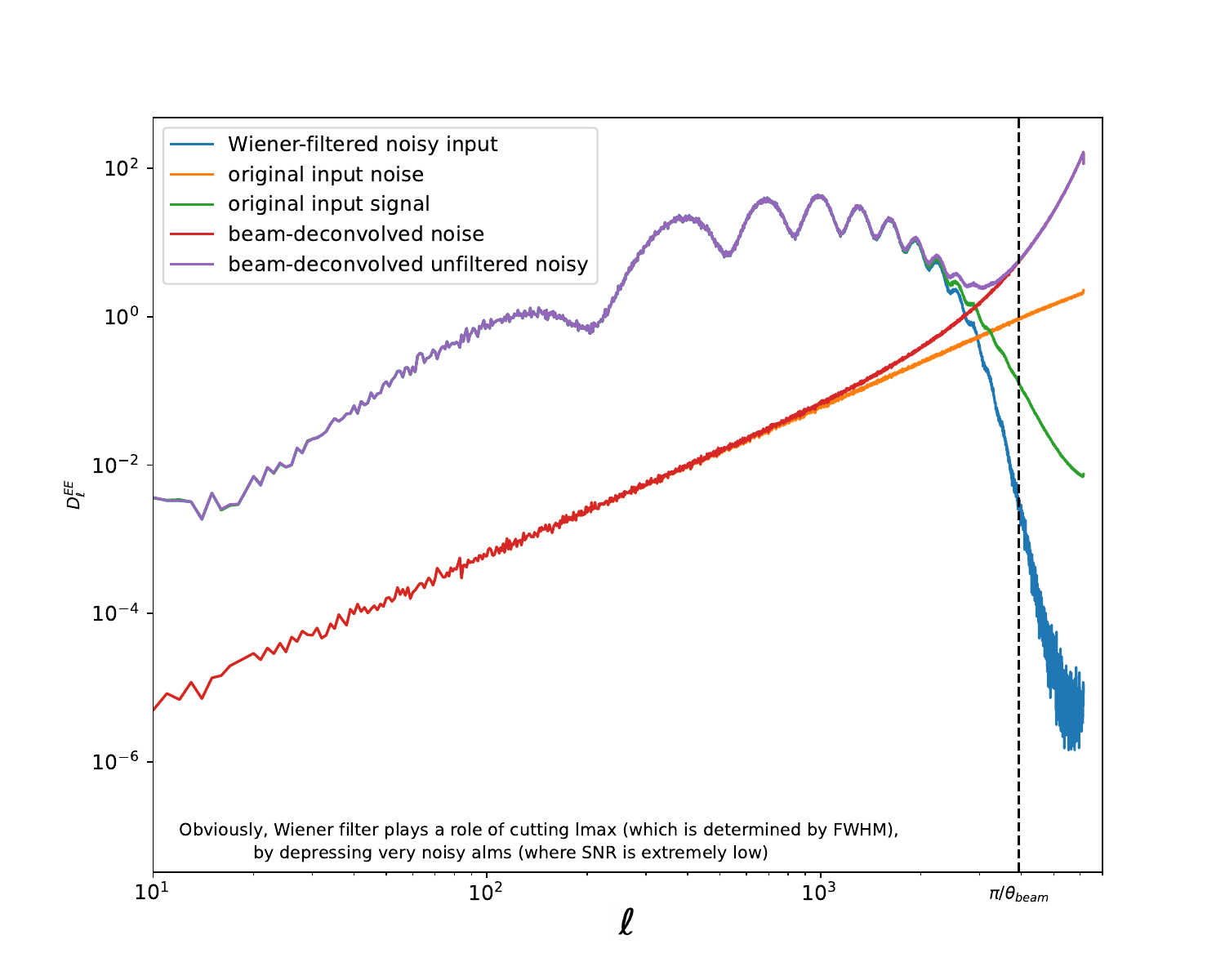}
	\caption{The input signal, noise, signal + noise, and filtered input. Here we show the power spectrum of a deconvolved field before and after Wiener filter
	One can see that the filtered input become suppressed where SNR is low ($\ell > 3000$), and the filtered input is almost the same as the signal where SNR is high.}
	\label{fig:Wiener_filter}
\end{figure}

\begin{figure}
	\centering
	\includegraphics[width=\columnwidth]{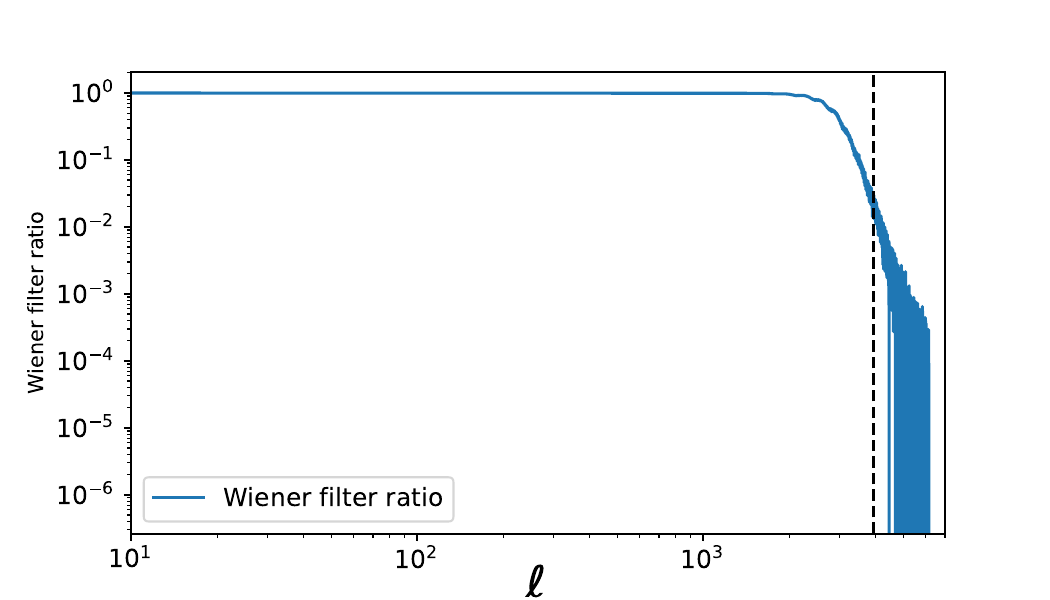}
	\caption{The Wiener filter ratio of the deconvolved field}
	\label{fig:Wiener_filter2}
\end{figure}

\begin{figure}
	\centering
	\includegraphics[width=\columnwidth]{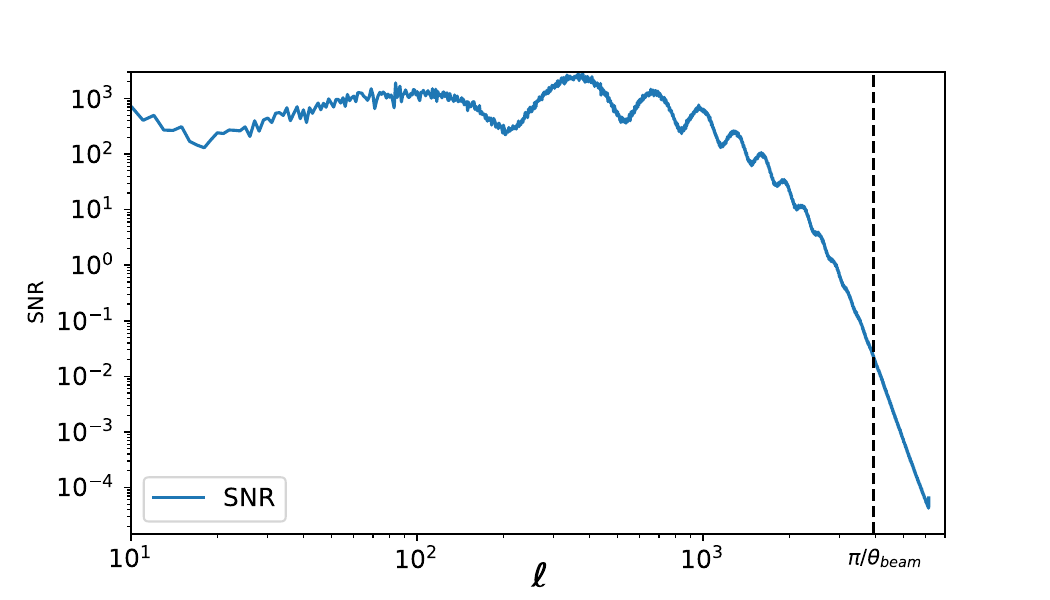}
	\caption{The signal-to-noise ratio of the deconvolved field}
	\label{fig:Wiener_filter3}
\end{figure}

\begin{figure}
	\centering
	\includegraphics[width=\columnwidth]{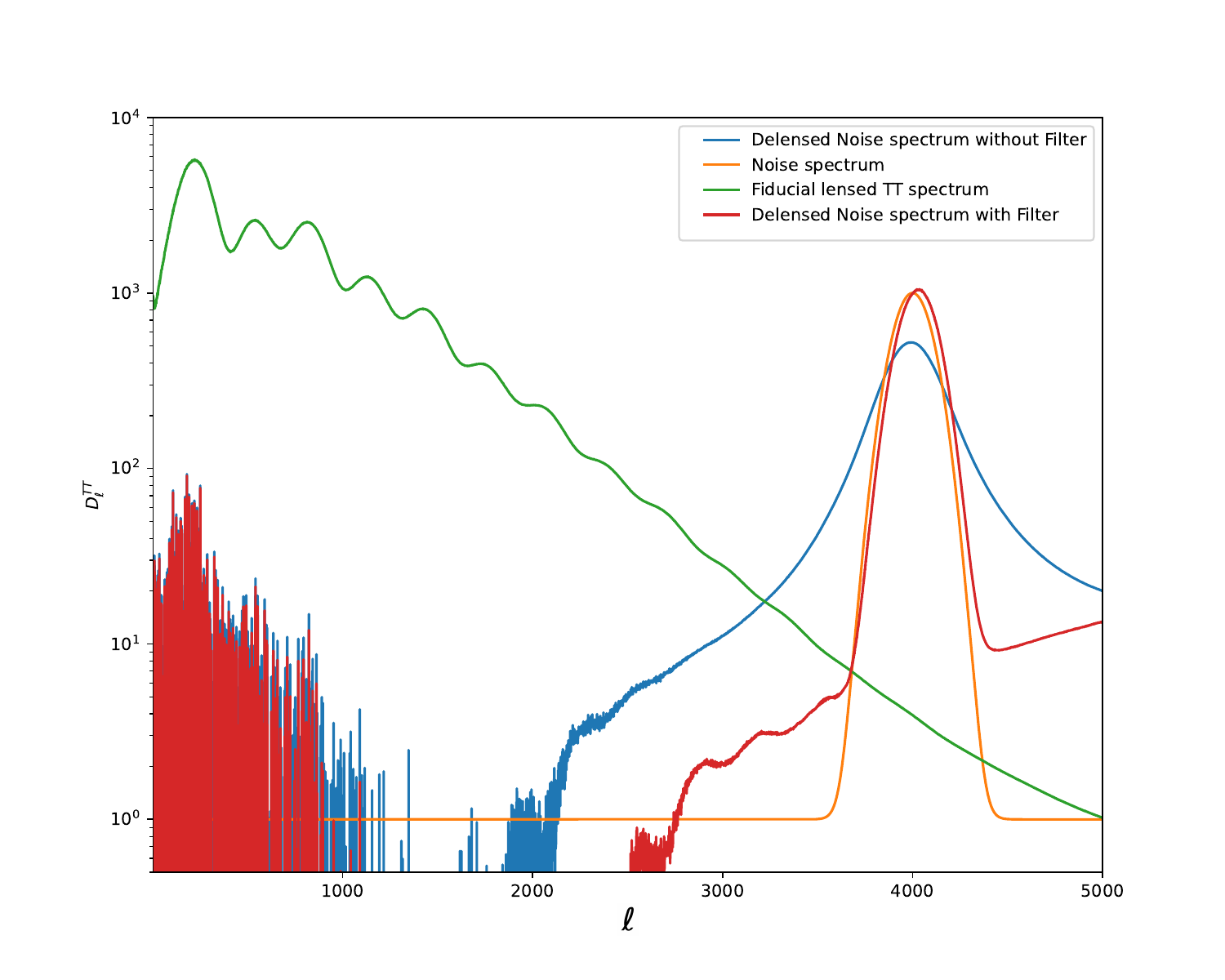}
	\caption{The lensed temperature power spectrum and noise curve in green and orange respectively. The noise power is given a Gaussian packet for illustration. The noise curve after delensing 
	is shown with and without filtering in red and blue respectively. One can see that the filter suppress the expansion of the noise power spectrum during delensing, and the shape of delensed noise curve change a little with a small shift.}
	\label{fig:filtered_noise_unchange}
\end{figure}

We here make a demonstration by two independent simulations. We show the influence on the input data in Fig.\ref{fig:Wiener_filter}, Fig.\ref{fig:Wiener_filter2} and Fig.\ref{fig:Wiener_filter3}. In addition, we artificially introduce a Gaussian peak in the noise power spectrum as a toy simulation to emphasize its impact on the delensing output in Fig.\ref{fig:filtered_noise_unchange}.
From Fig.\ref{fig:Wiener_filter}, Fig.\ref{fig:Wiener_filter2} and Fig.\ref{fig:Wiener_filter3} we can see that the filtered input become suppressed at where SNR is low, and the filtered input is almost unchanged at where SNR is high.
The Wiener filter become suppressed after $\ell \approx 2500$, and so does the noisy input (blue). 
In this way, the Wiener filter divide the noisiest modes $(\ell > 2500)$ from the clean modes $(\ell < 2500)$, avoiding inputting the very noisy modes into delensing process.

We know that lensing will smooth a peak in power spectrum, and when we delens an observed field which consists of lensed CMB and instrumental noise, we effectively lens them both.
For the lensed CMB, we expect this lensing will recover it to the primary CMB. However for the noise, we will smooth its power spectrum due to this lensing.
From Fig.\ref{fig:filtered_noise_unchange}, we see that the unfiltered noise curve exceeds the signal in locations that were signal dominated before delensing, which decreases the SNR of these modes. 
When delensing, the noise in a certain single mode will be remapped to other modes according to the inverse-lensing angle. This causes the noise power to spread, particularly in the case of inhomogeneous noise. So the delensed residual field will increase. 
Thanks to the filtering, which leaves the noise curves essentially unchanged, preventing the situation above from happening as much as possible.
In a word,  filtering improve the delensing by improving the input SNR.

Notice that there are still noise residuals after the filtering, which means that the lensing template or the remapped field consists of noise residuals as well.
We can perform noise debias on power spectrum by simulations as in our baseline.
Besides, as mentioned before, the filtering will suppress the signal more or less, and we can re-scale the lensing template to compensate for this suppression if necessary, by calculating the transfer function from the signal-only simulations.

\section{Method for adding lensing template (LT) as an independent channel}\label{app:l3}

In this method, we treat the lensing template as an additional pseudo-channel, instead of doing any delensing process on the map level, then calculate the auto and cross
power spectra to form the theoretical model vector. Our map channels will be $[B^{obs}, B^{LT}]$, then the dataset vector can be $\hat{X}_{\ell}=[C_{\ell}^{obs},C_{\ell}^{LT \times obs},C_{\ell}^{LT}]$. Each element of $\hat{X}_{\ell}$ is parameterized as: 

\begin{align}
    &C_{\ell}^{obs} = rC_{\ell}^{tens}(r=1) + A_LC_{\ell}^{lens} + N_{\ell}^{BB}, \\
    &C_{\ell}^{LT \times obs} = \sqrt{A_L}C_{\ell}^{lens},\\
    &C_{\ell}^{LT} = A_LC_{\ell}^{lens} + N_{\ell}^{temp}. 
\end{align}

where parameter $A_L$ scales the amplitude of the reconstructed lensing B-mode. The covariance matrix of $\hat{X}_{\ell}$ is estimated from 500 signal+noise simulations to capture all the complexities.

\section{Contribution of Tensor and Scalar components}\label{app:component}
We check the contribution of the polarization modes induced by the tensor and scalar fluctuation to the delensing results by simulation. The input maps share the same seed, and we adopt the current upper limit for $r$, which is $r=0.036$. We use two delensing methods to determine the contribution of the three component combination: tensor($r=0.036$) only, scalar only, tensor($r=0.036$) + scalar, to the delensed B-mode and the lensing B-mode template respectively.

\begin{figure}
	\centering
	\includegraphics[width=\columnwidth]{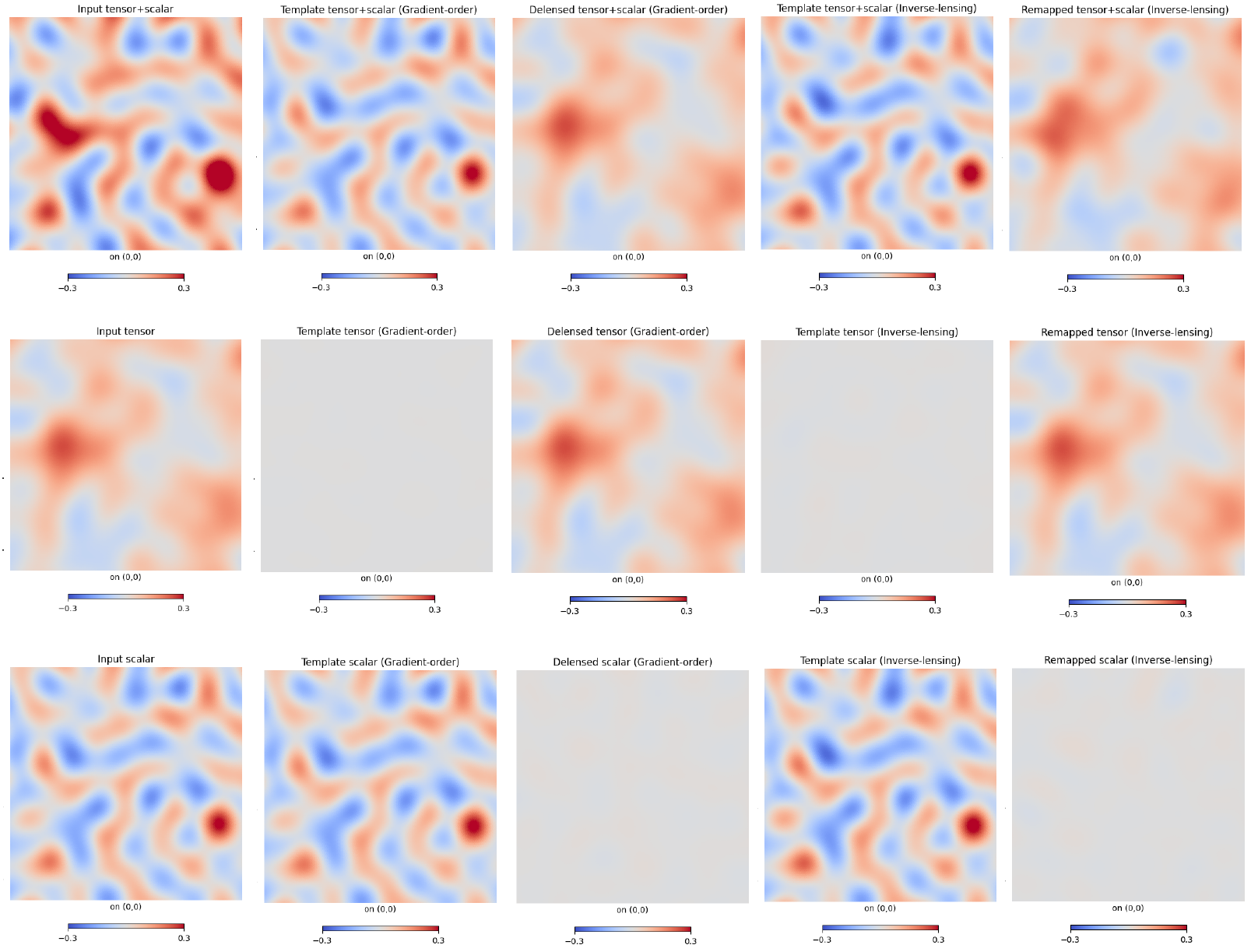}
	\caption{The input B map, the lensing template B map and the delensed B map for three components and two delensing methods.}
	\label{fig:tensor_scalar_map}
\end{figure}

An example figure of the maps are shown in Fig.\ref{fig:tensor_scalar_map}. For both the inverse-lensing method and the gradient-order method, the results are similar. Firstly, we can see that the tensor only case have negligible contribution to the total lensing B-mode template, which mostly comes from the scalar component. This is evident from the comparison between the first column and the second and the fourth columns. For the realistic observation (tensor + scalar case), we see that the lensing B-mode template is similar to the one from scalar only, this is precisely why we could write Eq.(\ref{EQ25}). Secondly, corresponding to the last point, we can see that the delensed B-mode of tensor only case remains almost unchanged compared to the input, while the delensed B-mode of the scalar only case fades evidently compared to the input. This is evident from the third and the fifth columns.  For the realistic observation (tensor + scalar case), we see that the delensed B-mode is similar to the input tensor B-mode, this is the reason that we gave Eq.(\ref{EQ28}).

\begin{figure}
	\centering
	\includegraphics[width=\columnwidth]{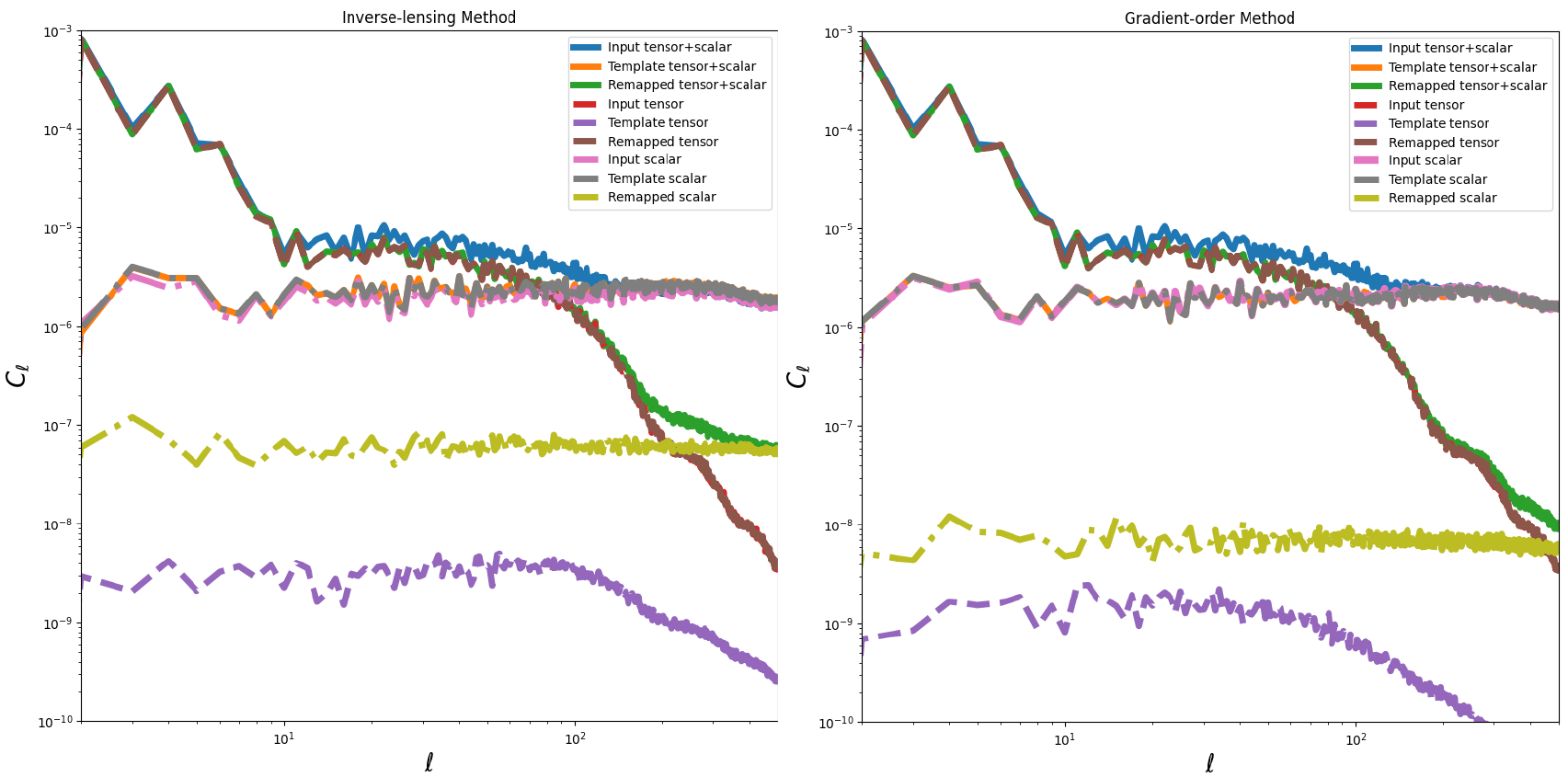}
	\caption{The input B map, the lensing template B map and the delensed B map for three components and two delensing methods.}
	\label{fig:tensor_scalar_cl}
\end{figure}

The corresponding power spectra of the simulation are shown in Fig.\ref{fig:tensor_scalar_cl}. It is evident that the power of the lensing B-mode template from tensor only case is negligible compared to the one of the input tensor B-mode and the one of the lensing B-mode template from scalar only case. Besides, the power of the delensed B-mode from tensor only case is very close to the one of the input tensor B-mode.
These confirm our viewpoints at the power spectrum level.

\acknowledgments

We thank Zuhui Fan, Jiakang Han, Zirui Zhang, Yi-Ming Wang, Zi-Xuan Zhang, for useful discussion. This study is supported by the National Key R\&D Program of China No.2020YFC2201601, National Natural Science Foundation of China No.12403005.
We acknowledge the use of many python packages: \texttt{CAMB} \cite{lewis2011camb}, \texttt{Healpy} \cite{zonca2019healpy}, \texttt{Lenspyx} \cite{carron2020lenspyx},  \texttt{CMBlensplus} \cite{namikawa2021CMBlensplus}, \texttt{NaMaster} \cite{alonso2023namaster} and \texttt{Cobaya} \cite{Torrado_2021}, \cite{torrado2020cobaya}.



\bibliographystyle{JHEP}
\bibliography{biblio.bib}


\end{document}